\documentclass[structabstract]{aa}

\usepackage{graphicx}
\usepackage{natbib}
\usepackage{morefloats}

\bibpunct{(}{)}{;}{a}{}{,} 
\usepackage{txfonts}
\begin{document}
 \title{A double radio relic in the merging galaxy cluster \object{ZwCl~0008.8+5215}} 

\titlerunning{A double radio relic in the galaxy cluster ZwCl~0008.8+5215}

   \author{R.~J. van Weeren\inst{1}
         \and M.~Hoeft \inst{2}
         \and H.~J.~A. R\"ottgering\inst{1}
         \and M.~Br\"uggen \inst{3}
         \and H.~T.~Intema \inst{4}
         \and S. van Velzen \inst{5}
          }

   \institute{Leiden Observatory, Leiden University,
              P.O. Box 9513, NL-2300 RA Leiden, The Netherlands\\
              \email{rvweeren@strw.leidenuniv.nl}
                \and Th\"uringer Landessternwarte Tautenburg, Sternwarte 5, 07778, Tautenburg, Germany
                 \and Jacobs University Bremen, P.O. Box 750561, 28725 Bremen, Germany
                 \and National Radio Astronomy Observatory, 520 Edgemont Road, Charlottesville, VA 22903-2475, USA
                 \and Department of Astrophysics, Institute for Mathematics, Astrophysics and Particle Physics, Radboud University, PO~Box~9010, 6500~GL Nijmegen, The Netherlands
                 }


 
\abstract
   {Some merging galaxy clusters host diffuse elongated radio sources, also called radio relics. It is proposed that these radio relics trace shock waves in the intracluster medium (ICM) created during a cluster merger event. Within the shock waves particles are accelerated to relativistic energies, and in the presence of a magnetic field synchrotron radiation will be emitted. Here we present Giant Metrewave Radio Telescope (GMRT) and Westerbork Synthesis Radio Telescope (WSRT) observations of a new double relic in the galaxy cluster ZwCl~0008.8+5215.}
   {The aim of the observation is to understand the phenomenon of radio relics. 
   }
   {We carried out radio continuum observations at 241 and 610~MHz with the GMRT, and $1.3-1.8$~GHz observations with the WSRT in full polarization mode. Optical V, R, and I band images of the cluster were taken with the 2.5m Isaac Newton Telescope (INT). An optical spectrum, to determine the redshift of the cluster, was taken with the William Herschel Telescope (WHT).
      }
   {Our observations show the presence of a double radio relic in the galaxy cluster  ZwCl~0008.8+5215, for which we find a spectroscopic redshift of $z = 0.1032 \pm 0.0018$ from an optical spectrum of one of the cD galaxies.  The spectral index of the two relics steepens inwards to the cluster center. For part of the relics, we measure a polarization fraction in the range ${\sim 5-25\%}$. 
   A ROSAT X-ray image displays an elongated ICM and the large-scale distribution of galaxies reveals two cluster cores, all pointing towards a binary cluster merger event. The radio relics are located symmetrically with respect to the X-ray center of the cluster, along the proposed merger axis. The relics have a linear extent of 1.4~Mpc and 290~kpc. This factor of five difference in linear size is unlike that of previously known double relic systems, for which the sizes do not differ by more than a factor of two.  
   }
{ We conclude that the double relics in ZwCl~0008.8+5215 are best explained by two outward moving shock waves in which particles are (re)accelerated trough the diffusive shock acceleration (DSA) mechanism. 
}
   \maketitle

\section{Introduction}
Radio relics are filamentary structures often located in the 
periphery of galaxy clusters. It is proposed that large 
radio relics trace shock waves generated by cluster 
merger events \citep{1998A&A...332..395E, 2001ApJ...562..233M}. 
At the shock front particles from the thermal gas are accelerated to relativistic energies
by DSA mechanism in a first-order Fermi process 
\citep{1977DoSSR.234R1306K, 1977ICRC...11..132A, 1978MNRAS.182..147B, 1978MNRAS.182..443B, 1978ApJ...221L..29B, 1983RPPh...46..973D, 1987PhR...154....1B, 1991SSRv...58..259J, 2001RPPh...64..429M}. In the presence of a magnetic field these particles emit synchrotron radiation at radio wavelengths. Another possibility mentioned by \cite{2005ApJ...627..733M}, is that the shock re-accelerates relativistic fossil electrons injected previously into the ICM by for example AGN. They note that from an observational point of view, this case will probably be indistinguishable from the direct shock acceleration mentioned above. 
An alternative scenario has been recently proposed, namely that relics arise from emission of secondary cosmic ray electrons \citep{2010arXiv1011.0729K}.

Some smaller radio relics ($\lesssim 500$kpc) have been 
been explained by old (``fossil'') radio plasma from a 
previous episode of AGN activity. These sources are 
called \emph{AGN relics} \citep[see][for an overview of 
the classification of diffuse radio sources]{2004rcfg.proc..335K}. 
The fossil radio plasma could also have been compressed, creating 
a radio \emph{phoenix} \citep{2001A&A...366...26E, 2002MNRAS.331.1011E}. 
Both radio phoenices and AGN relics, are characterized by a 
very steep ($\alpha \lesssim -1.5$, {$F_{\nu} \propto \nu^{\alpha}$, where $\alpha$ 
is the spectral index) and curved radio spectra  due to synchrotron 
and Inverse Compton (IC) losses. 

In the hierarchical model of structure formation galaxy 
cluster grow by the accretion of gas from the surrounding 
intergalactic medium (IGM) and through mergers with other 
clusters and galaxy groups. 
Large radio relics are exclusively found in disturbed 
clusters, indicative of merger activity. This supports the idea that shocks 
generated during cluster merger events can be responsible 
for the non-thermal radio emission. Hydrodynamical models 
of structure formation, including particle acceleration 
mechanisms \citep[e.g.,][]{2007MNRAS.375...77H, 2008MNRAS.391.1511H, 2008MNRAS.385.1242P, 2009MNRAS.393.1073B} 
make predictions about the location, orientation and 
radio power of relics in merging clusters. Amongst the 
several dozen radio relics known to date {\citep[e.g., ][]{2009A&A...508...75V, 2009ApJ...697.1341R, 2009A&A...507..639P, 2008A&A...486..347G, 2008A&A...489...69B, 2001ApJ...548..639K, 1999NewA....4..141G, 1991A&A...252..528G},  there are a 
few rare double relic systems, with two relics located 
symmetrically on opposite sites of the cluster center 
\citep[e.g., ][]{2010Sci...330..347V, 2010arXiv1011.4985B, 2009A&A...494..429B, 2009A&A...506.1083V, 2007A&A...463..937V, 2006Sci...314..791B, 1997MNRAS.290..577R}. 
These double relics can be used to constrain 
the merger geometry and timescales involved 
\citep{1999ApJ...518..603R}. If radio relics 
trace outward traveling shock waves in which DSA 
takes place, then the radio plasma in the post-shock 
region should have a steeper spectrum due to IC and 
synchrotron losses. 
For a relic seen close to edge-on,  
the luminosity profile across the width of the relic 
can then be used to constrain the magnetic fields 
strength at the location of the shock \citep[][]{2005ApJ...627..733M, 2010arXiv1004.2331F}. 
This is because 
the downstream luminosity profile should directly 
reflect the synchrotron losses, which in turn depend 
on the magnetic field strength.  
In \cite{2010Sci...330..347V} we presented observations of a  
double relic in the merging cluster \object{CIZA~J2242.8+5301}, which provided evidence for DSA in galaxy cluster merger shocks. For the largest  
relic we derived a magnetic field strength of about 5--7~$\mu$Gauss by modeling 
the relic's luminosity profile across the width of the relic (although a strength of about 1.2~$\mu$Gauss could not be completely ruled out).

Because there are only a few double relics systems known, 
we carried out an extensive search in the 1.4~GHz NVSS 
\citep{1998AJ....115.1693C}, 325~MHz WENSS 
\citep{1997A&AS..124..259R}, and 74~MHz VLSS 
\citep{2007AJ....134.1245C} surveys for the 
presence of arc-like radio structures around 
X-ray selected galaxy clusters from the ROSAT 
All-Sky Survey. This search already resulted 
in the discovery of several new radio relics, 
see \cite{2009A&A...506.1083V, 2009A&A...508...75V, 2009A&A...505..991V}. 
In this paper we present the discovery of a 
double relic in the galaxy cluster \object{ZwCl~0008.8+5215} 
which showed faint elongated structures 
in both NVSS and WENSS images.

The layout of this paper is as follows. 
In Sect.~\ref{sec:obs-reduction} we give 
an overview of the observations and the data 
reduction. In Sect.~\ref{sec:results} we present 
the radio and spectral index maps as well as 
optical images around the radio sources. 
In Sect.~\ref{sec:discussion} we discuss 
the merger scenario and the magnetic field 
strength at the location of the relics. We 
end with conclusions in Sect.~\ref{sec:conclusion}.

Throughout this paper we assume a $\Lambda$CDM 
cosmology with $H_{0} = 71$~km~s$^{-1}$~Mpc$^{-1}$, $\Omega_{m} = 0.3$, and $\Omega_{\Lambda} = 0.7$. 
All images are in the J2000 coordinate system.

\section{Observations \& Data Reduction}
\label{sec:obs-reduction}

\subsection{GMRT observations}
The GMRT observations were taken in dual-frequency mode, recording RR polarization at 610~MHz and LL polarization at 241~MHz. Total useable bandwidth was 30~MHz at 610~MHz and 6~MHz at 241~MHz. The GMRT software backend \citep[GSB; ][]{2009arXiv0910.1517R} was used giving 512 spectral channels. The total on source time was 220~min (3.7~hr). A summary of the observations is given in Table~\ref{tab:gmrtobservations}. 

For the reduction we used the NRAO Astronomical Image Processing System (AIPS) package. The 610~MHz dataset was visually inspected for the presence of radio frequency interference (RFI) which was subsequently removed. Strong RFI on short baselines was present in the 241~MHz band. This RFI was fitted and subtracted from the data using the technique described by \cite{2009ApJ...696..885A} which was implemented in  Obit \citep{2008PASP..120..439C}. Summarized, the fringe-stopped correlator output of a baseline oscillates with the fringe-stop period in the presence of RFI. The fitting routine tries to fit such a signal and subtracts it from the data \citep{2009ApJ...696..885A}. This has the advantage that most of the visibilities from the short baselines are not removed (i.e., flagged) and hence spatial structure on large angular scales is preserved. We assumed the RFI to be constant over periods of 10 min and required a minimal amplitude of 5~Jy for the RFI to be subtracted.  Further visual inspection of the data was carried out to remove some remaining RFI which could not be fitted because of short timescale variations of the RFI signal.

Amplitude and phase corrections were 
obtained for the calibrator sources 
using 5 neighboring channels free of RFI. 
These solutions were applied to the data  before determining the bandpass.
The bandpass was then applied and gain 
solutions for the full channel range were determined.
The fluxes of the primary calibrators 
were set according to the \cite{perleyandtaylor} 
extension to the \cite{1977A&A....61...99B} scale. 
Several rounds of phase self-calibration 
and two final rounds of amplitude and phase 
self-calibration were carried out. We used the 
polyhedron method \citep{1989ASPC....6..259P, 1992A&A...261..353C} 
for making the images to minimize the effects 
of non-coplanar baselines. Both in the 610 and 
241~MHz images, there were several bright sources 
in the field that limited the dynamic range. 
Direction dependent gain solutions for these 
sources were obtained and these sources were 
subtracted from the data. This method is commonly 
referred to as ``peeling'' \citep[e.g.,][]{2004SPIE.5489..817N}.

Final images were made using ``briggs'' weighting 
\citep[with robust set to 0.5,][]{briggs_phd}. Images 
were cleaned down to $2$ times the rms noise level 
($2\sigma_{\mathrm{rms}}$) within the clean boxes. 
The images were corrected for the primary beam 
response\footnote{http://gmrt.ncra.tifr.res.in/gmrt\_hpage/Users/doc/manual/
 
 UsersManual/node27.html}. The uncertainty in the calibration 
of the absolute flux-scale is between $5-10\%$, see \cite{2004ApJ...612..974C}. 

\begin{table}
\begin{center}
\caption{GMRT observations}
\begin{tabular}{lll}
\hline
\hline
& 241~MHz  & 610~MHz  \\
\hline
Observation date &   November 22, 2009&November 22, 2009 \\
Usable bandwidth & 6~MHz & 30~MHz\\
Channel width &62.5~kHz & 62.5~kHz\\
Polarization & LL & RR \\
Integration time & 8 sec & 8 sec\\
Total on-source time & 220 min & 220 min\\
Beam size                                        & $14.65\arcsec \times 12.5\arcsec$ & $6.3\arcsec \times 5.3\arcsec$ \\
Rms noise ($\sigma_{\rm{rms}}$) & 483~$\mu$Jy~beam$^{-1}$ &   38~$\mu$Jy~beam$^{-1}$ \\
\hline
\end{tabular}
\label{tab:gmrtobservations}
\end{center}
\end{table}

\subsection{WSRT $1.3-1.7$~GHz observations}

\begin{table*}
\begin{center}
\caption{WSRT observations}
\begin{tabular}{ll}
\hline
\hline
Frequency bands 21 cm (IFs)             & 1311, 1330, 1350, 1370, 1392, 1410, 1432, 1450~MHz  \\
Frequency bands 18 cm (IFs)            &  1650, 1668, 1686, 1704, 1722, 1740, 1758, 1776~MHz\\
Bandwidth per IF                                 & 20~MHz \\
Number of channels per IF                     & 64 \\
Channel width                                      & 312.5~kHz\\
Polarization			                 & XX, YY, XY, XY \\
Observation dates		                &  28 March, 25 \& 31 May,  7 \& 16 June 7, 2009 \\
Integration time                                & 30~sec \\
Total on-source time		             & $12.5$~hr (21cm) , $12.5$~hr (18cm)\\
Beam size                                        & $23.5\arcsec \times 17.0\arcsec$~(21cm), $23.5\arcsec \times 17.0\arcsec$~(18cm) \\
Rms noise ($\sigma_{\rm{rms}}$) & 27~$\mu$Jy~beam$^{-1}$~(21cm),   33~$\mu$Jy~beam$^{-1}$~(18cm)\\
\hline
\hline
\end{tabular}
\label{tab:wsrtobservations}
\end{center}
\end{table*}

WSRT observations of ZwCl~0008.8+5215 were taken in the L-band, see Table \ref{tab:wsrtobservations}. The observations were spread out over various runs of several hours, resulting in more or less full 12 hours synthesis coverage. Frequency switching, between the 18 and 21~cm setups, was employed every 5 min to increase the spectral baseline of the observations. Each setup had a total bandwidth of  160~MHz, evenly divided over 8 sidebands (IF). All four polarization products were recored with 64 channels per IF. 

The data were calibrated using the CASA\footnote{http://casa.nrao.edu/} package. The L-band receivers of the WSRT antennas have linearly polarized feeds\footnote{The WSRT records $\rm{XX}=\rm{I}-\rm{Q}$, $\rm{YY}=\rm{I}-\rm{Q}$, $\rm{XY} = -\rm{U }+ i\rm{V}$, and $\rm{XY} = -\rm{U}-i\rm{V}$. I, Q, U, and V are the standard Stokes parameters}. Bandpass and gain solutions were determined from observations of the standard calibrators 3C48, 3C138, 3C147, 3C286, and CTD93. Time ranges for antennas affected by shadowing were removed.  Some IFs were also affected by RFI and had to be partly flagged. The fluxes for the calibrators were set according to the \cite{perleyandtaylor} extension to the \cite{1977A&A....61...99B} scale. Frequency (channel) dependent leakage terms (D-terms) were calculated from observations of a bright unpolarized point source, while the polarization angles were set using either 3C138 or 3C286. We assumed $-66.0\degr$ and $15.0\degr$ for the R-L phase difference of 3C286 and 3C138, respectively. The data were then exported into AIPS and several rounds of phase self-calibration, followed by two rounds of amplitude and phase self-calibration, were carried out. The solutions for the amplitude and phase self-calibration were determined combining both XX and YY polarizations because Stokes Q is not necessarily zero. Separate images for each IF were made (16 in total) using robust weighting of 0.75 (21~cm) and 1.25 (18~cm) giving roughly the same resolution. These images were corrected for the primary beam attenuation ($A(r)$)
\begin{equation}
A(r) = \cos^{6}{\left(r \times \nu \times C\right)} \mbox{ ,}
\end{equation}
with $r$ the distance from the pointing center in degrees, $\nu$ the observing frequency in GHz and $C=68$, a constant\footnote{from the WSRT Guide to Observations}. 

Images were cleaned to about $2\sigma_{\mathrm{rms}}$ and clean boxes were used. For each frequency setup the eight images from each IF were combined, convolving the images to the same resolution of $23.5\arcsec \times 17.0\arcsec$ and weighting the images inversely proportional to $\sigma^{2}_{\mathrm{rms}}$. The images are centered at frequencies of 1382 and 1714~MHz. 
The final noise levels are 27 and 33~$\mu$Jy~beam$^{-1}$ for the 21~cm and 18~cm images, respectively.

\subsection{Optical images \& spectroscopy}
Optical images of the cluster were made with the Wide Field 
Camera (WFC) on the INT. The observation were taken between 
1 and 8 October, 2009. The seeing varied between 
0.9 and 1.3\arcsec~during 
the observations. Most nights were photometric. 
Total integration time was about 6000~sec per filter. The data were reduced in a 
standard way using IRAF 
\citep{1986SPIE..627..733T, 1993ASPC...52..173T} 
and the \emph{mscred} package \citep{1998ASPC..145...53V}. 
The R and I band images were fringe corrected. 
The individual exposures were averaged, pixels 
were rejected above $3\sigma_{\mathrm{rms}}$ 
to remove cosmic rays and other artifacts.
Observations of standard stars on photometric 
nights were used to calibrate the flux scale. 
The images have a depth (signal to noise radio 
(SNR) of 5 for point sources) of approximately 
$24.5$, $23.9$, $23.3$ magnitude (Vega) in the 
V, R and I band, respectively. 

To determine the redshift of the cluster a 600~sec WHT ACAM long-slit spectrum, with the V400 grating, was taken of \object{2MASX~J00112171+5231437} \citep{2003yCat.7233....0S}. On the INT images this galaxy was identified being the largest cD galaxy of the cluster. The spectrum was taken on November 2, 2010 with a slit width of 1.5\arcsec. Standard long-slit calibration was done in IRAF.

\section{Results}
\label{sec:results}

\subsection{Redshift of ZwCl~0008.8+5215}
\label{sec:spectrum}
 No redshift is available in the literature for the galaxy cluster ZwCl~0008.8+5215, with coordinates RA~00$^{\rm{h}}$11$^{\rm{m}}$25.6$^{\rm{s}}$, DEC~+52\degr 31\arcmin41\arcsec. The cluster is located relatively close to the galactic plane at a galactic latitude $b= -9.86\degr$. Galactic extinction is 0.812 mag in the R-band and 0.111 in the K-band according to \cite{1998ApJ...500..525S}.  The spectrum of cD galaxy \object{2MASX~J00112171+5231437} is shown in Fig.~\ref{fig:spectra}. From this spectrum we determine a redshift of $0.1032 \pm 0.018$ for the galaxy, which we adopt as the redshift for ZwCl~0008.8+5215. 2MASX~J00112171+5231437 is associated with a complex disturbed radio source, see Sect.~\ref{sec:agn}.

\begin{figure}
\begin{center}
\includegraphics[angle =90, trim =0cm 0cm 0cm 0cm,width=0.5\textwidth]{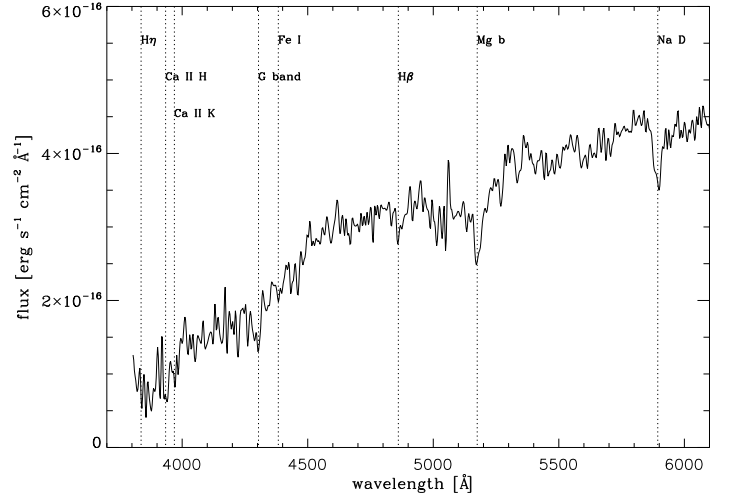}
\end{center}
\caption{Spectrum of the galaxy 2MASX~J00112171+5231437 taken with the WHT ACAM V400 grating. Various absorption features are indicated.}
\label{fig:spectra}
\end{figure}

\subsection{Thermal ICM and galaxy distribution}
\label{sec:xray}
ZwCl~0008.8+5215 is seen in the ROSAT All-Sky Survey as an 
east-west elongated source,  see Fig.~\ref{fig:xray},  
and listed as \object{1RXS~J001145.3+523147} \citep{1999A&A...349..389V}. 
Using the redshift and the ROSAT count rate we find an X-ray 
luminosity ($L_{\rm{X,~0.1-2.4~keV}}$) of $\sim 5 \times 10^{43}$~erg~s$^{-1}$. 
With the $L_{\rm{X,~0.1-2.4~keV}}$-temperature scaling 
relation from \cite{2009A&A...498..361P}, Table B.2 {\it BCES Orthogonal Fitting Method}, 
we find a corresponding temperature of $\sim 3 \mbox{ to } 4$~keV.

We computed galaxy iso-densities from the INT images. We first created 
a catalog of objects using Sextractor \citep{1996A&AS..117..393B}. 
We then removed all point-like objects (i.e., stars) from the catalogs. 
To exclude galaxies not belonging to the cluster we selected only galaxies 
with R$-$I and V$-$R colors within $0.15$ magnitude from the average color 
of the massive elliptical cD galaxy 2MASX~J00112171+5231437 (see Fig.~\ref{fig:SEF} and Sect.~\ref{sec:spectrum}). The 
range of $0.15$ in the colors was taken to maximize the contrast of the 
cluster with respect to the fore and background galaxies in the field, 
but not being too restrictive so that a sufficient number of candidate 
cluster members was selected. The galaxy iso-density contours are shown 
in Fig.~\ref{fig:xray}. The cluster shows a pronounced bimodal structure,  
with two cores separated by about 700~kpc. The cluster extends somewhat 
further in the east-west direction than the X-ray emission from ROSAT.  The cD galaxy  2MASX~J00112171+5231437 belongs to the western subcluster (i.e., it is located at the center of the subcluster). 
The eastern subcluster also hosts a separate cD galaxy (\object{2MASX~J00121892+5233460}, see Fig.~\ref{fig:SD}). Although we do not have a spectroscopic redshift for this galaxy, the (i) color and (ii) R and K magnitudes are in agreement with a subcluster located at the same redshift as the western subcluster \citep[e.g.,][]{2003MNRAS.339..173W, 2007A&A...464..879D}. The same is true for the other massive elliptical galaxies found in both subclusters, see Sect.~\ref{sec:agn}. Therefore, both the X-ray and optical observations point towards a bi-modal galaxy cluster, indicative of an ongoing merger event. As we will show in the next sections, the radio observations also point towards a merger scenario. 

\begin{figure}
\begin{center}
\includegraphics[angle =90, trim =0cm 0cm 0cm 0cm,width=0.5\textwidth]{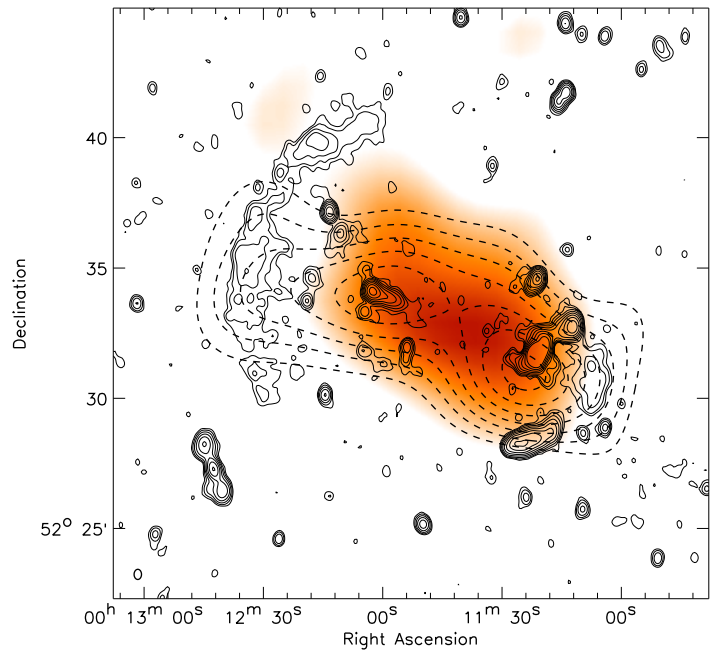}
\end{center}
\caption{X-ray emission from ROSAT, tracing the thermal ICM, is shown by the color image. The original image from the ROSAT All Sky Survey was convolved with a 200\arcsec~FWHM Gaussian. Solid contours are from the WSRT 1382~MHz image and drawn at levels of ${[1, 2, 4, 8, \ldots]} \times 4\sigma_{\mathrm{rms}}$. The resolution of the WSRT image is $23.5\arcsec \times 17.0\arcsec$. Dashed contours show the galaxy iso-density distribution derived from the INT images. Contours are drawn at ${[1, 1.1, 1.2, 1.3, 1.4 \ldots]}  \times 0.38$ galaxies arcmin$^{-2}$ using the color cuts as described in Sect.~\ref{sec:xray}. }
\label{fig:xray}
\end{figure}

\subsection{Radio continuum maps}
\begin{table*}
\begin{center}
\caption{Relic \& source properties}
\begin{tabular}{llllllll}
\hline
\hline
Source & $S_{241\mbox{ } \rm{MHz}}$ & $S_{610\mbox{ } \rm{MHz}}$ & $S_{1382 \mbox{ } \rm{MHz}}$ & $S_{1714 \mbox{ } \rm{MHz}}$ & $\alpha_{241 \rm{MHz}}^{1714 \mbox{ } \rm{MHz}}$& $P_{1.4\mbox{ }\rm{GHz}}$&LLS$^{b}$ \\
 &    Jy & mJy&mJy & mJy & & $10^{24}$ W~Hz$^{-1}$& kpc\\
\hline
RW & $0.11 \pm 0.03$ & $56 \pm 8$ & $11\pm1.2$  & $8.9 \pm 1.2$ & $-1.49\pm 0.12^{a}$ &  0.37 & 290\\  
RE & $0.82 \pm 0.09$ & $230 \pm 25$ & $56\pm 3.5$ & $37 \pm 2.7$& $-1.59\pm0.06^{a}$  & 1.8&1400\\  
A   & $0.186 \pm 0.020 $ & $86 \pm 9.0 $ & $37 \pm 1.9$ & $ 29 \pm 1.6$ & $-0.96 \pm 0.06$ &1.1 & 270\\
B &  $0.112\pm 0.015  $  & $87 \pm 9.0 $ & $ 34\pm 1.9 $ & $ 28 \pm1.6$ & $-0.81\pm 0.06$ &  0.99&289\\
C & $0.423 \pm 0.045 $  &  $205 \pm 21$ & $ 88 \pm 4.6$ & $ 69 \pm 3.6 $ & $-0.94 \pm 0.06$  & 2.6&249 \\
D & \ldots &  $2.2 \pm 0.3$ & $1.06 \pm 0.1 $ & $1.2 \pm 0.1$ &  $-0.63 \pm 0.16^{c}$ &0.030 &\ldots \\
E & $2.81 \pm 0.28 $ & $1016 \pm 102 $ & $387 \pm 20$ & $302 \pm 15$ &$-1.14\pm0.05$&11.9 &200\\
F & $0.280 \pm 0.030$ & $135\pm14$& $58 \pm3$&$47\pm3$&  $-0.93 \pm 0.06$ &1.7 &151\\
G & $0.040 \pm 0.005$ & $16\pm2 $ &  $ 7.3 \pm 0.4$ & $6.3\pm0.4$ &  $-0.94 \pm 0.07$ & 0.22& 38\\
H & $\ldots $ & $8.7\pm1.2$ & $2.6\pm0.2$& $2.7\pm0.2$& $-1.18 \pm 0.15^{c}$ &0.081 &60\\
I      & \ldots & $2.8 \pm 0.5$ & $1.2 \pm 0.19$ & $0.76 \pm 0.18$  &   $-1.15 \pm 0.23^{c}$ &0.037 & 45\\
\hline
\hline
\end{tabular}
\label{tab:relicflux}
\end{center}
$^{a}$ see also Fig.~\ref{fig:relicflux}\\
$^{b}$ largest linear size\\
$^{c}$ between 610 and 1714~MHz
\end{table*}

\begin{figure}
\begin{center}
\includegraphics[angle =90, trim =0cm 0cm 0cm 0cm,width=0.5\textwidth]{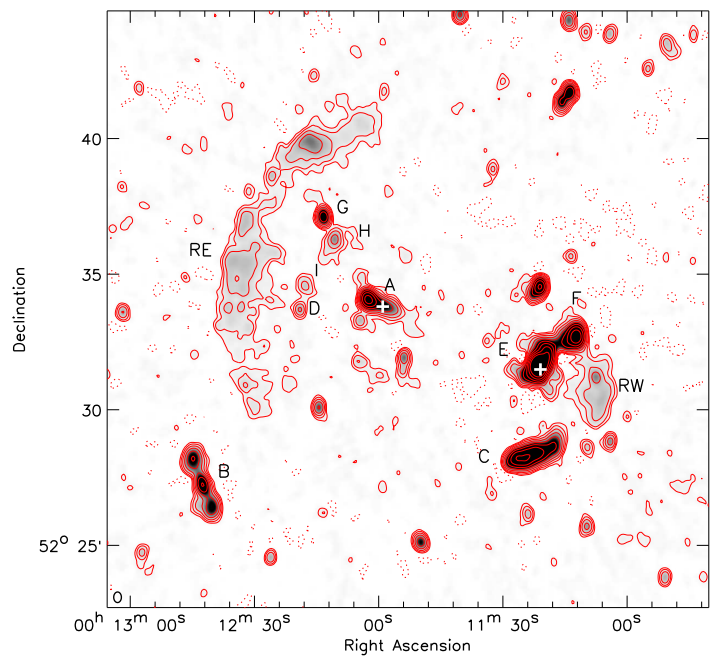}
\end{center}
\caption{WSRT 1382~MHz image. Contour levels are drawn at ${[-1, 1, 2, 4, 8, \ldots]} \times 4\sigma_{\mathrm{rms}}$. Negative contours are shown by the dotted lines. The beam size is $23.5\arcsec \times 17.0\arcsec$ and shown in the bottom left corner of the image. Sources are labeled as in Fig.~\ref{fig:gmrt610_labels}. The white~+~symbols mark the centers of the two subclusters based on the iso-density contours from Fig.~\ref{fig:xray}. }
\label{fig:wsrt21cm}
\end{figure}

\begin{figure}
\begin{center}
\includegraphics[angle =90, trim =0cm 0cm 0cm 0cm,width=0.5\textwidth]{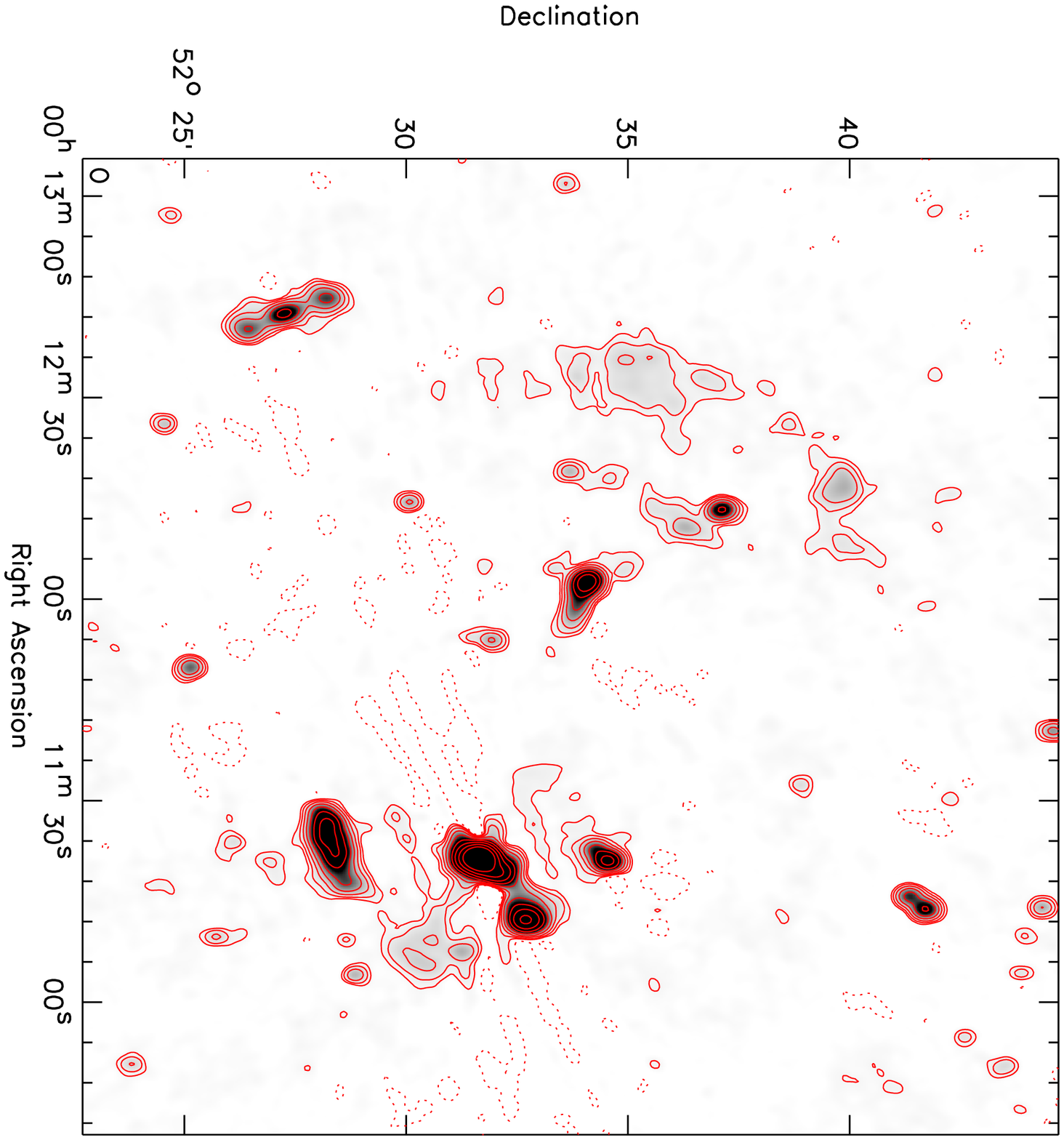}
\end{center}
\caption{WSRT 1714~MHz image. Contour levels are drawn at ${[-1, 1, 2, 4, 8, \ldots]} \times 4\sigma_{\mathrm{rms}}$.  Negative contours are shown by the dotted lines. The beam size is $23.5\arcsec \times 17.0\arcsec$ and shown in the bottom left corner of the image.}
\label{fig:wsrt18cm}
\end{figure}

\begin{figure*}
\begin{center}
\includegraphics[angle =90, trim =0cm 0cm 0cm 0cm,width=1.0\textwidth]{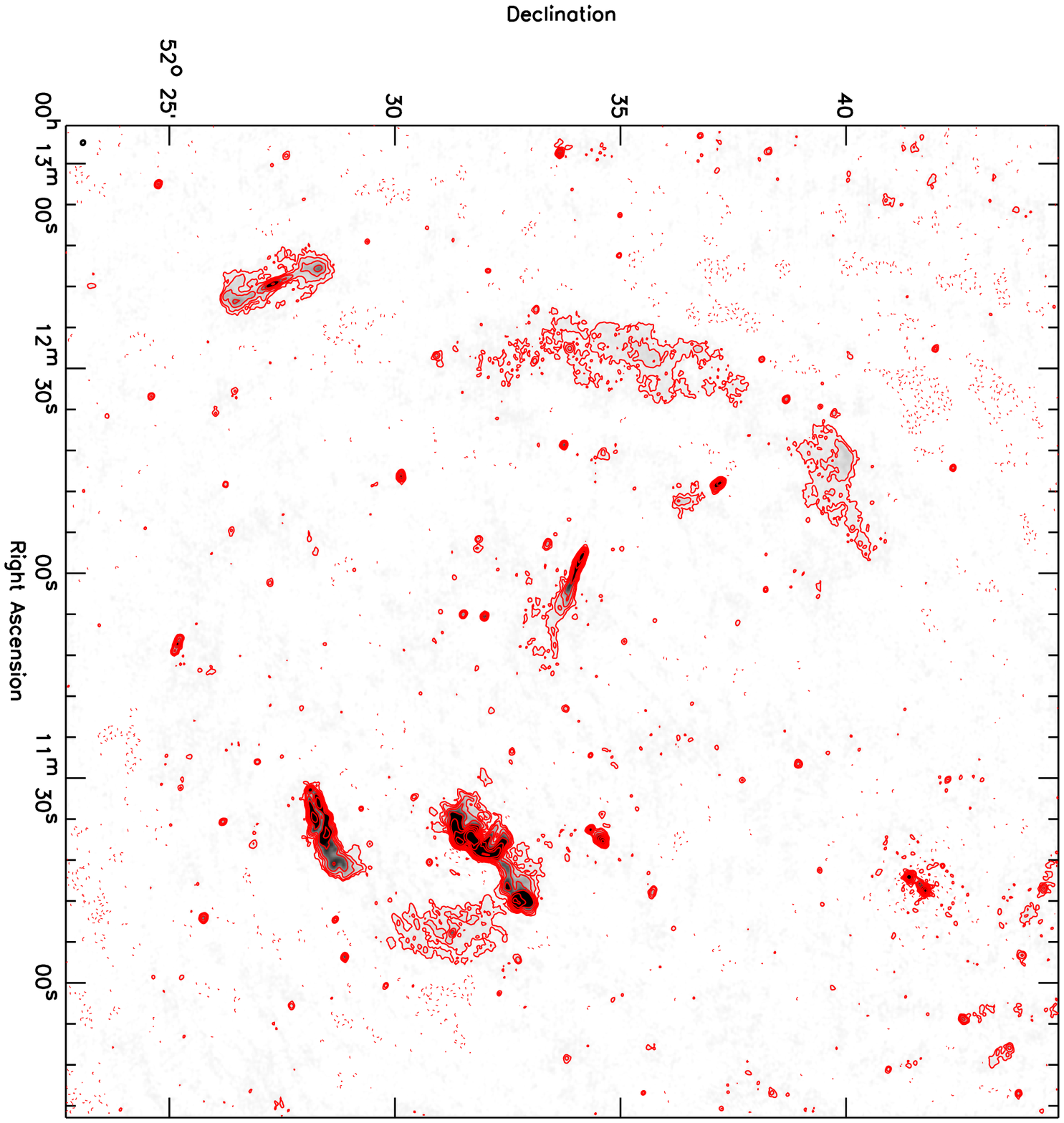}
\end{center}
\caption{GMRT 610~MHz image. Contour levels are drawn at ${[-1, 1, 2, 4, 8, \ldots]} \times 4\sigma_{\mathrm{rms}}$.  Negative contours are shown by the dotted lines. The beam size is $6.3\arcsec \times 5.3\arcsec$ and shown in the bottom left corner of the image.}
\label{fig:gmrt610}
\end{figure*}

\begin{figure}
\begin{center}
\includegraphics[angle =90, trim =0cm 0cm 0cm 0cm,width=0.5\textwidth]{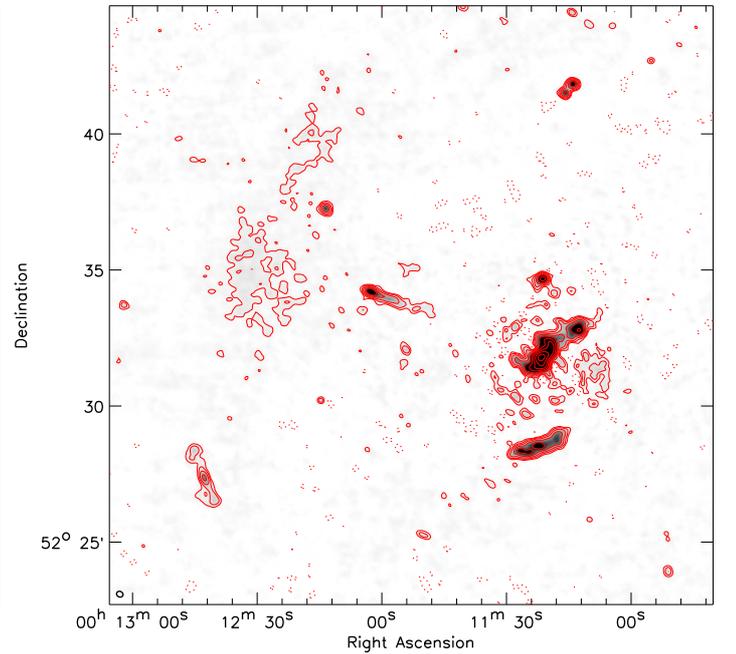}
\end{center}
\caption{GMRT 241 MHz image. Contour levels are drawn at $[-1, 1, 2, 4, 8, \ldots] \times 4\sigma_{\mathrm{rms}}$.  Negative contours are shown by the dotted lines. The beam size is $14.65\arcsec \times 12.50\arcsec$ and shown in the bottom left corner of the image.}
\label{fig:gmrt241}
\end{figure}

The WSRT 1382~MHz image is shown in Fig.~\ref{fig:wsrt21cm}. 
It reveals a large arc of diffuse emission on the east side 
of the cluster and a smaller faint diffuse source on the west 
side of the cluster, symmetrically with respect to the cluster 
center. We classify these sources as radio relics based on their 
location with respect to the cluster center, their morphology, 
and the lack of optical counterparts. The relics are located about 850~kpc from the center of the X-ray emission. 
Several complex tailed 
radio sources, related to AGN activity, are also visible. 
The WSRT 1714~MHz image is similar to the 1382~MHz image, 
although the overall signal to noise ratio is less, 
therefore revealing less of the diffuse extended relics. 
The radio relics are also visible in the GMRT 610 and 
241~MHz images (Figs.~\ref{fig:gmrt610} and \ref{fig:gmrt241}), 
although at 241 MHz the SNR on the relics is $\lesssim 5$ per beam.

To facilitate the discussion we have labeled various 
sources in Figs.~\ref{fig:wsrt21cm} and \ref{fig:gmrt610_labels}. Optical 
overlays can be found Sect.~\ref{sec:agn}. 
The integrated fluxes, spectral indices, radio power 
and largest linear size for the two relics (RE \& RW) 
are displayed in Table~\ref{tab:relicflux}.

Relic RE consist of two parts, a smaller region of 
emission to the north and a larger one in the south 
(most clearly seen in Figs.~\ref{fig:gmrt610} and 
\ref{fig:gmrt610_labels}). In the 1382~MHz image 
the two regions are seen connected. The eastern boundary 
of RE is somewhat more pronounced, while on the western 
side the emission fades more slowly in the direction 
of the cluster center. The relic has a total extent 
of 1.4~Mpc. The surface brightness varies across the 
relic fading at the extreme northern and southern ends. 
The northern diffuse patch has a ``notch'' like region 
of higher surface brightness. Relic RW has a much smaller extent of 290~kpc. 
The western boundary is more pronounced in the WSRT 
images. A compact source in the middle of RW is 
associated with a background galaxy.

The cluster also hosts a number of complex radio 
sources related to AGN activity, for a short 
discussion on these sources see Sect.~\ref{sec:agn}.  

\begin{figure}
\begin{center}
\includegraphics[angle =90, trim =0cm 0cm 0cm 0cm,width=0.5\textwidth]{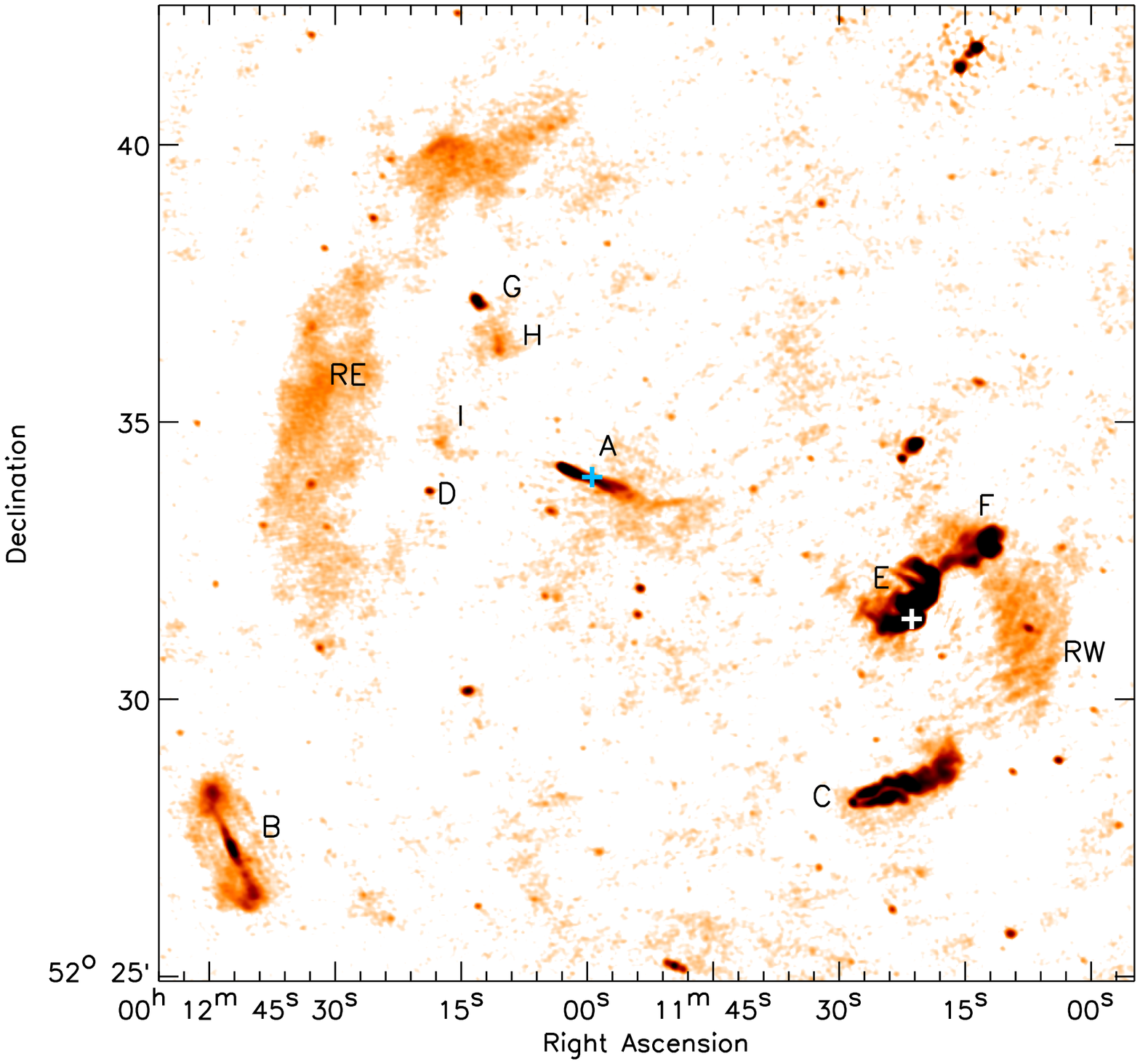}
\end{center}
\caption{GMRT 610 MHz image with sources labeled. The blue and white~+~symbols mark  
the centers of the two subclusters based on the iso-density contours from Fig.~\ref{fig:xray}. }
\label{fig:gmrt610_labels}
\end{figure}

\subsection{Spectral index and polarization maps}
\label{sec:spixmap}
A spectral index map was computed using both the WSRT and GMRT images, including only common UV ranges.   
Both the WSRT and GMRT datasets have relatively good inner 
UV-coverage. The largest detectable angular scale is limited 
to about 16\arcmin~at 610~MHz, which is sufficient not to 
resolve out the extended radio relics. The inclusion of 
maps at four different frequencies enables us to map the 
spectral index over the low surface brightness radio relics. 
Spectral index maps made with only two frequency images were too 
noisy to map the spectral index across the relics. 
The spectral index map was created by fitting a single power-law 
through the flux measurements at 241, 610, 1382, and 1714~MHz. 
In this way, we only fitted for the slope and normalization of the radio spectrum, ensuring the 
number of free variables in the fit remains as low as possible (at the cost of detecting spectral curvature). The technique of combining maps at more than two frequencies has another advantage that errors in the maps arising from RFI, calibration errors, deconvolution errors, slightly different UV coverage, etc., are suppressed in the spectral index map as long as they do not correlate at the same location and spatial frequencies on the sky.

Pixels in the spectral index map were blanked if any of 
corresponding pixels in the individual maps fell below
 $1.5\sigma_{\rm{rms}}$. Special care was taken about 
the precise alignment of the maps, we slightly shifted 
the GMRT maps by about  a quarter of the synthesized beam, removing a 
small spectral index gradient visible across all the point 
sources. The result is shown in Fig.~\ref{fig:spix}. 
 
For relic RW, the spectral index steepens to the north 
and eastwards to the cluster center, from $-0.9$ to $-2.0$. 
The spectral index for relic RE also varies roughly 
between $-0.9$ and $-2.0$. The overall spectral index 
at the east side of relic RE is about $-1.2$ There 
is an overall trend of spectral steepening towards 
the west, see also Fig.~\ref{fig:relicprofile}. The 
spectral index is correlated with the surface brightness 
of the relic, the brightest parts have a flatter spectral index (see also Fig.~\ref{fig:relicprofile}).

\begin{figure*}
\begin{center}
\includegraphics[angle =90, trim =0cm 0cm 0cm 0cm,width=0.49\textwidth]{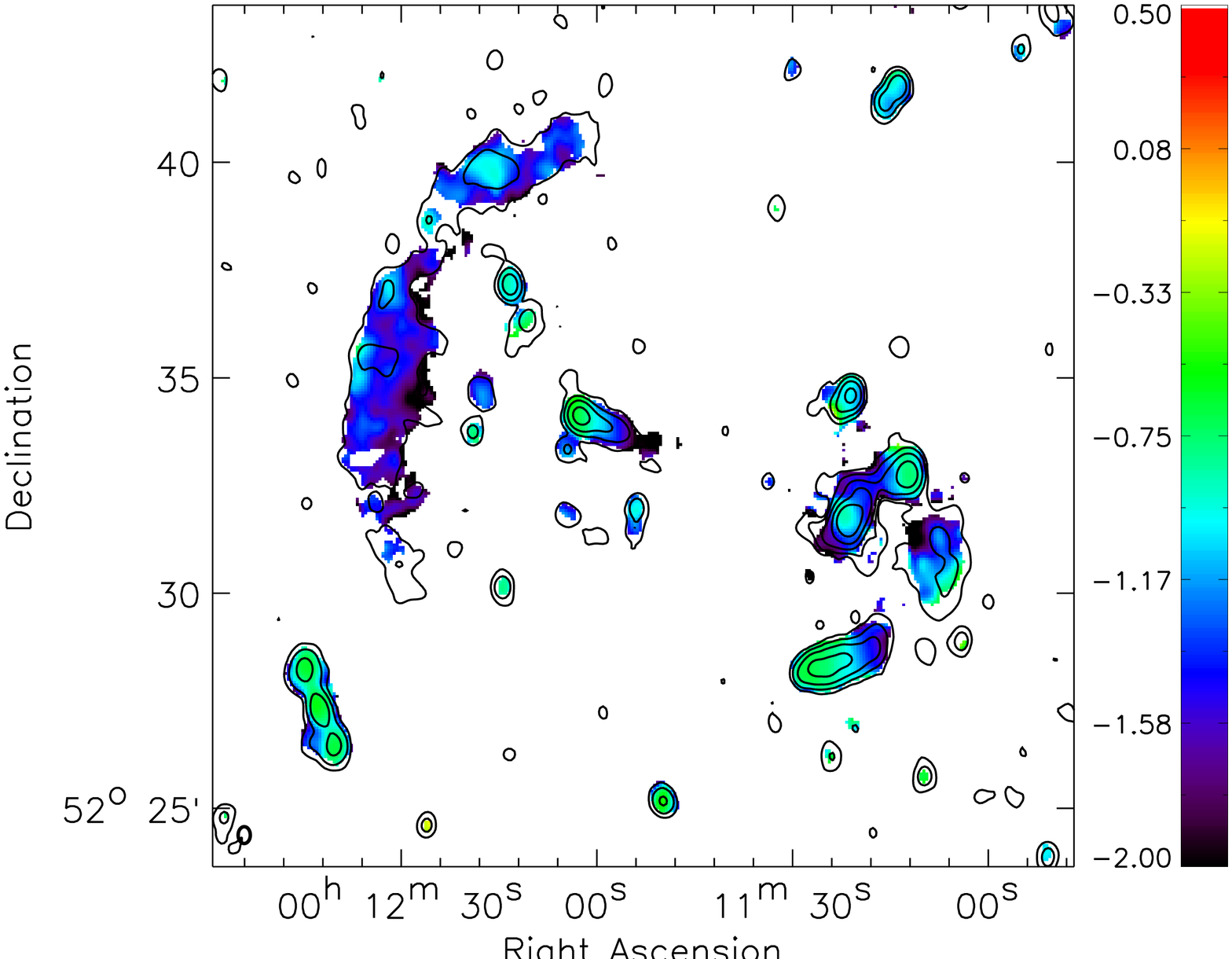}
\includegraphics[angle =90, trim =0cm 0cm 0cm 0cm,width=0.49\textwidth]{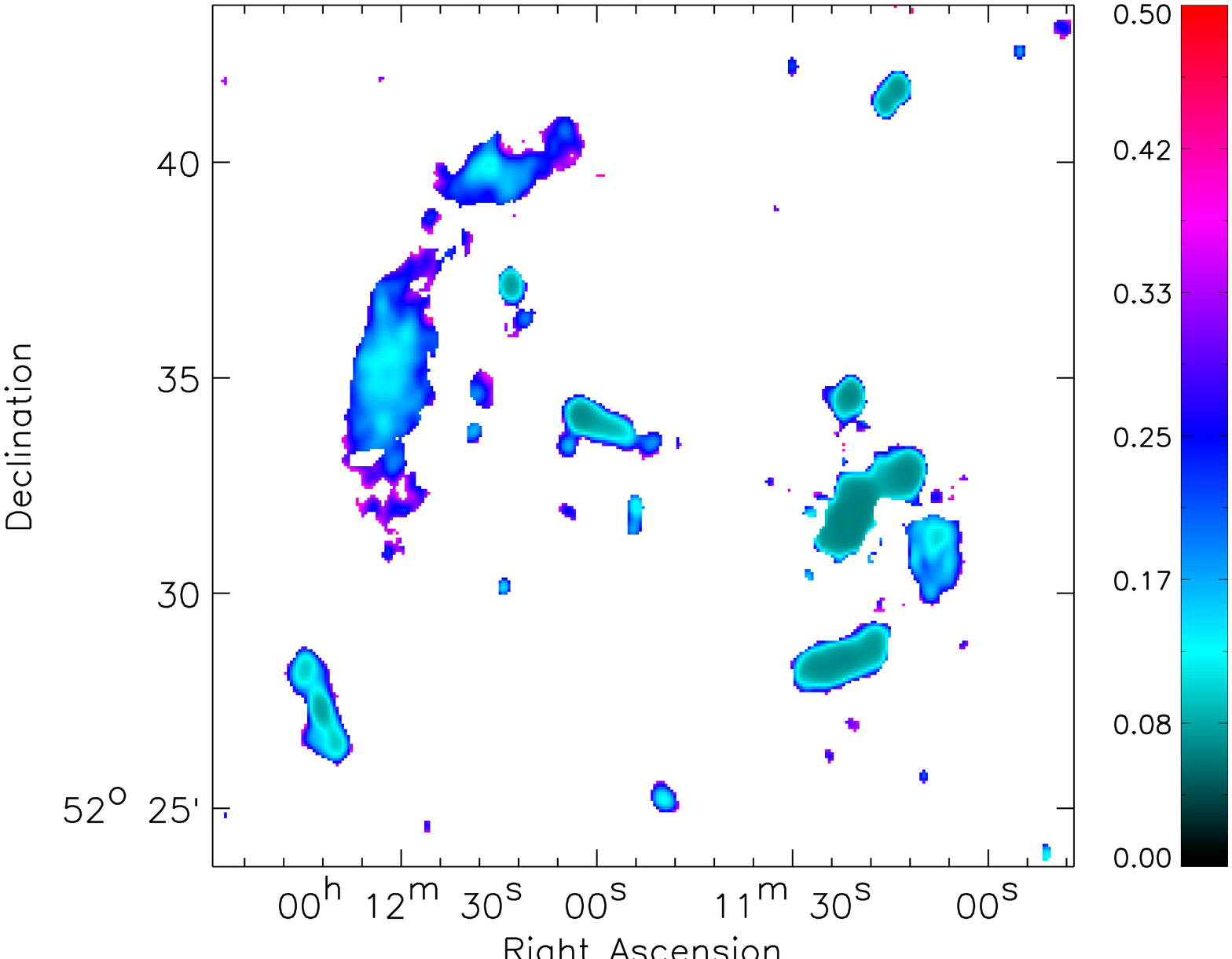}
\end{center}
\caption{Left: Spectral index map. Spectral indices were 
computed by fitting a single power-law radio spectrum 
through the fluxes at 241, 610, 1382, and 1714~MHz. Solid contours 
are from the 1382~MHz WSRT map and drawn at 
levels of ${[1, 4, 16, 64, \ldots]} \times 6\sigma_{\mathrm{rms}}$. 
The resolution of the map is $23.5\arcsec \times 17.0\arcsec$. 
Right: Spectral index uncertainty map. The map is computed on the 
basis of $\sigma_{\mathrm{rms}}$ values for the individual maps. }
\label{fig:spix}
\end{figure*}

The polarization map from the WSRT at 1382~MHz is 
shown in Fig.~\ref{fig:polzwcl52}. No useful polarization 
information could be extracted from the WSRT 18~cm observations. 
The polarization map reveals that most of the compact sources  
are polarized below the 5\% level. Some polarized emission is 
detected from the two radio relics, although at low SNR. 
For relic RW the polarization fraction is around $5-10\%$ \citep[reported polarization fractions are 
corrected for Ricean bias;][]{1974ApJ...194..249W}. For 
RE the polarization fraction varies, with a maximum of $\sim 25\%$. For 
the fainter parts of the relics no polarized emission is detected, 
but this is expected if the relics are polarized at the 30\% level or less. 
Most polarization E-vectors are aligned perpendicular to the major 
axis of the two relics (for the parts where polarized emission was detected).

Spectral index and polarization properties for the compact sources are discussed in Sect.~\ref{sec:agn}.

\begin{figure}
\begin{center}
\includegraphics[angle =90, trim =0cm 0cm 0cm 0cm,width=0.5\textwidth]{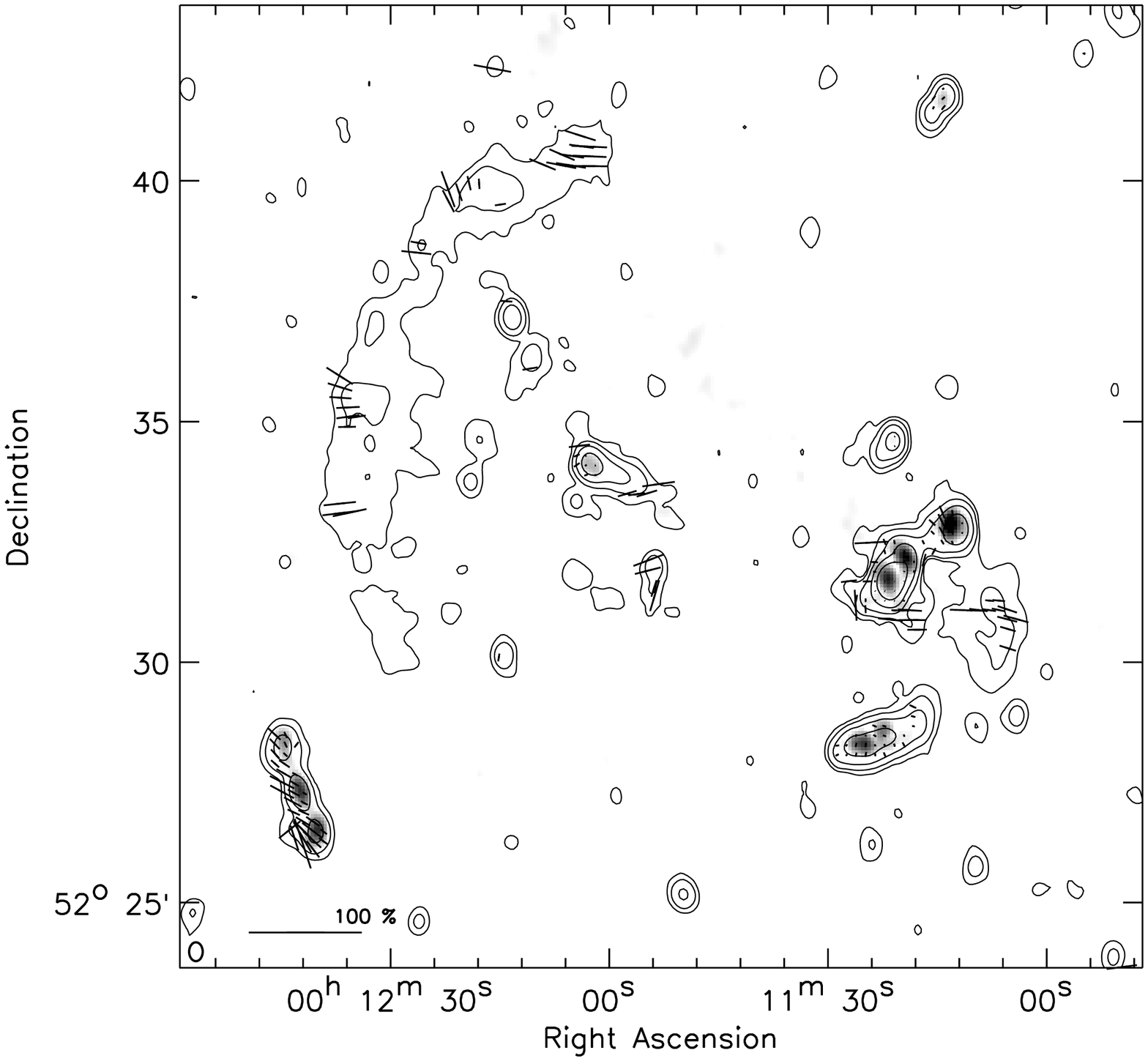}
\end{center}
\caption{WSRT 1382~MHz polarization E-vector map. Total polarized intensity 
is shown as grayscale image.  Vectors depict the polarization E-vectors, 
their length represents the polarization fraction. The length of the E-vectors are corrected 
for Ricean bias \citep{1974ApJ...194..249W}. 
A reference vector for a polarization fraction of 100\% is shown in 
the bottom left corner. No vectors were drawn for pixels with a 
SNR $< 5$ in the total polarized intensity image. 
Contour levels are drawn at ${[1, 2, 4, 8, \ldots]} \times 5\sigma_{\mathrm{rms}}$ 
and are from the Stokes I 1382~MHz image. The beam size 
is $23.5\arcsec \times 17.0\arcsec$ and shown in the bottom left corner of the image.}
\label{fig:polzwcl52}
\end{figure}

\subsection{Radio galaxies in the cluster}

The cluster hosts several interesting tailed radio sources. 
Radio overlays on optical images are shown 
in Figs.~\ref{fig:SA} to \ref{fig:SEF}. Sources are referred 
to as in Fig.~\ref{fig:gmrt610_labels}. The morphology of these 
radio sources is as expected for a system undergoing a merger, with high galaxy velocities 
with respect to the ICM. The radio powers reported for the sources in Table~\ref{tab:relicflux}, are consistent with them being {\rm FR-I} sources \citep{1974MNRAS.167P..31F} located in the cluster \citep[e.g.,][]{1991MNRAS.249..164O, 1994ASPC...54..319O,1995ApJ...451...88B}. 

Source~A is a ``head-tail'' source belonging to the 
elliptical galaxy \object{2MASX~J00120320+5234132}. The spectral 
index steepens along the tail, from $-0.6$ to $-2.1$ to the west. 
The ``head'' is polarized at the 1\% level. 
The tail has a total extent of 270~kpc 
in the 610~MHz image.  
Source~B is a double-lobe radio source 
(also listed as \object{NVSS J001242+522717} or \object{B0010.0+5210}) 
consisting of a central core and two radio lobes. Additional diffuse 
emission is seen surrounding the lobes and the central radio core. 
The spectral index does not vary much over the core and radio lobes, 
the average spectral index is about $-0.6$. The radio emission from 
the source has a polarization 
fraction between 10 and 20\%. An optical 
counterpart (\object{2MASX~J00124217+5227182}) is centered 
on the radio core. The source could be a cluster 
member as its color is similar to that of other galaxies in the cluster.

Source~C is a narrow angle tailed 
(NAT) radio source with \object{2MASX~J00112859+5228096} being the 
optical counterpart. The spectral index steepens 
from $-0.6$ to $-1.7$ along the tails as expected for a radio 
galaxy moving eastwards with respect to the ICM. 
The radio emission is polarized at the $2-4\%$ level. 
Source~D belongs to the cD galaxy \object{2MASX~J00121892+5233460} in the eastern part of the cluster and has a spectral index of $-0.8$. 
No polarized emission is detected from the source.

Source~E is the brightest radio source (\object{4C~+52.01}) 
in the cluster and has a very complex morphology. 
The radio emission has a sharp western boundary, 
while the eastern boundary is more diffuse with ``fans'' of emission. 
The counterpart is the galaxy \object{2MASX~J00112171+5231437}, 
which has a close companion to the west. The spectral 
index steepens to the north and south from $-0.9$ to $-1.7$. 
Fractional polarization for the brightest emission is below 1\%.

Source~F has a complex morphology and is 
located to the northwest of E. There is a 
bridge of steep spectrum emission (with $\alpha \sim-1.4$) between 
E and F. The polarization fraction of F varies 
between $2$ and $10\%$. The brighter parts of the source 
have a spectral index of $-0.75$. Because the spectral 
index here is flatter than the bridge of emission between E and F, 
this suggests the source is a separate radio galaxy with the 
emission not coming from E. The most likely counterpart 
is the galaxy \object{2MASS~J00111135+5232421}. The complex morphology 
of both E and F suggest that the radio sources are  
significantly disturbed, possibly due to the merger event. 
The radio morphology of both E and F suggest the sources 
moves westwards with respect to the ICM.

We could not identify optical counterparts for 
sources~G, H and I. However, a diffraction spike in the INT images 
from a nearby bright star partly covers source~G, likely blocking our 
view of the optical counterpart. The high surface brightness of 
G indicates the source is probably associated with an AGN.

\label{sec:agn}
\begin{figure}
\begin{center}
\includegraphics[angle =90, trim =0cm 0cm 0cm 0cm,width=0.5\textwidth]{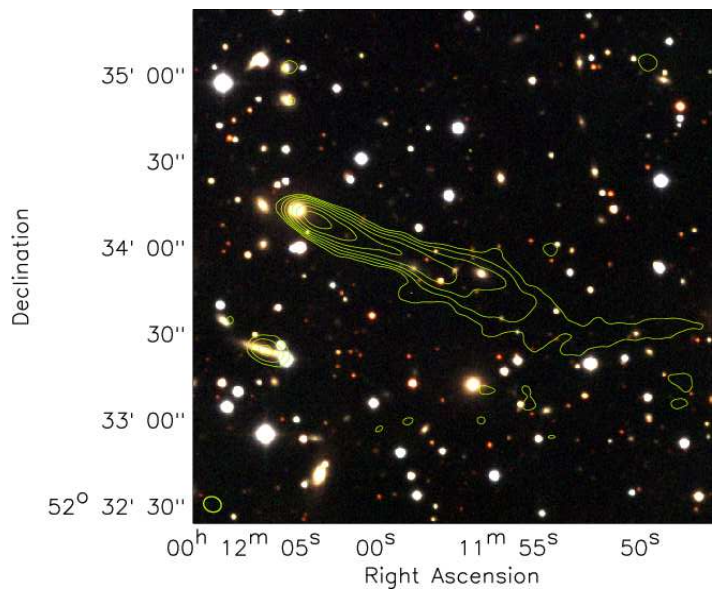}
\end{center}
\caption{Source A. GMRT 610~MHz contour are drawn at levels of $\sqrt{[1, 2, 4, 8, \ldots]} \times 4\sigma_{\mathrm{rms}}$. The beam size is $11.75\arcsec \times 7.65\arcsec$ and shown in the bottom left corner of the image.}
\label{fig:SA}
\end{figure}

\begin{figure}
\begin{center}
\includegraphics[angle =90, trim =0cm 0cm 0cm 0cm,width=0.5\textwidth]{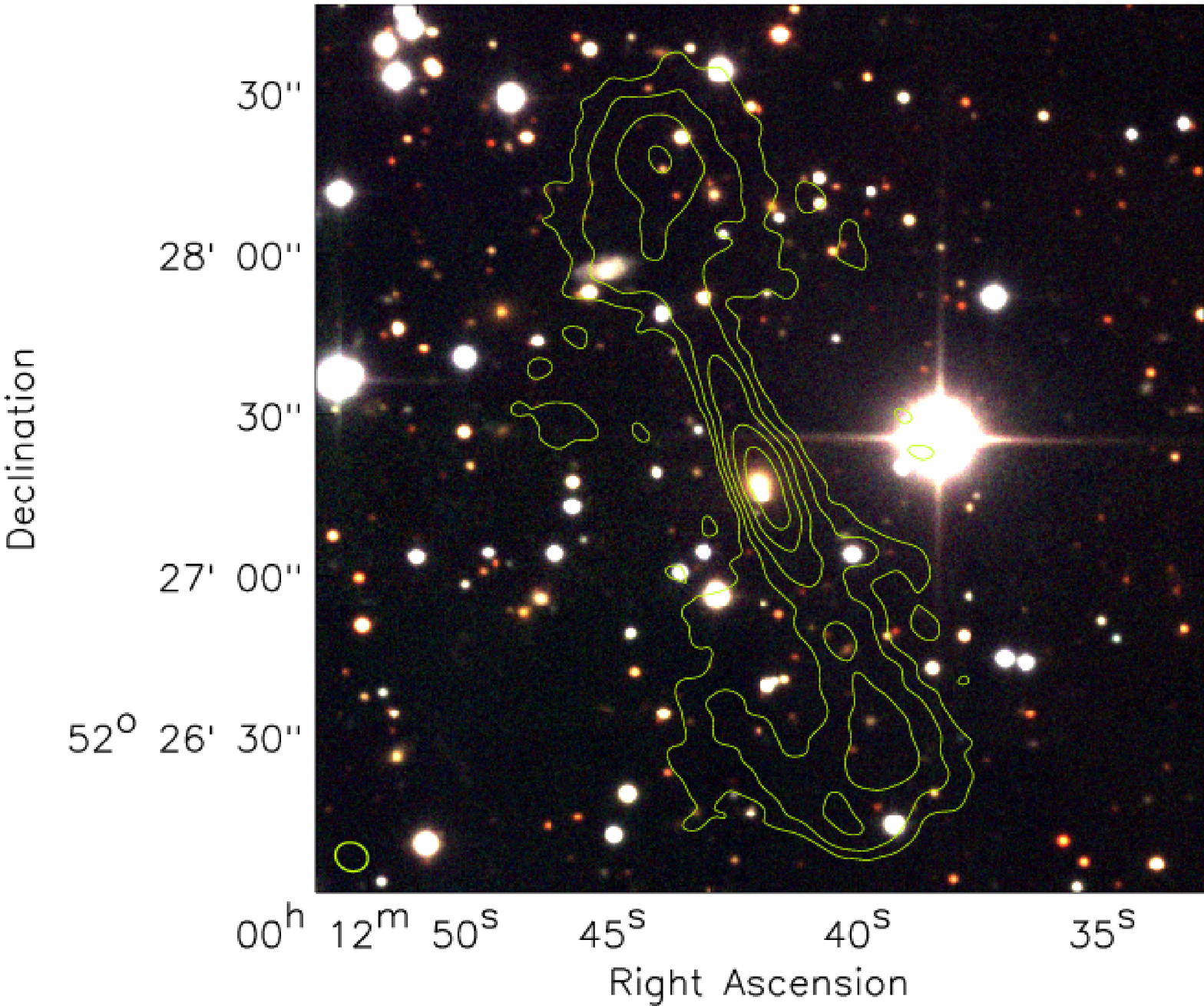}
\end{center}
\caption{Source B. GMRT 610~MHz contour levels are displayed as in Fig.~\ref{fig:SA}.}
\label{fig:SB}
\end{figure}

\begin{figure}
\begin{center}
\includegraphics[angle =90, trim =0cm 0cm 0cm 0cm,width=0.5\textwidth]{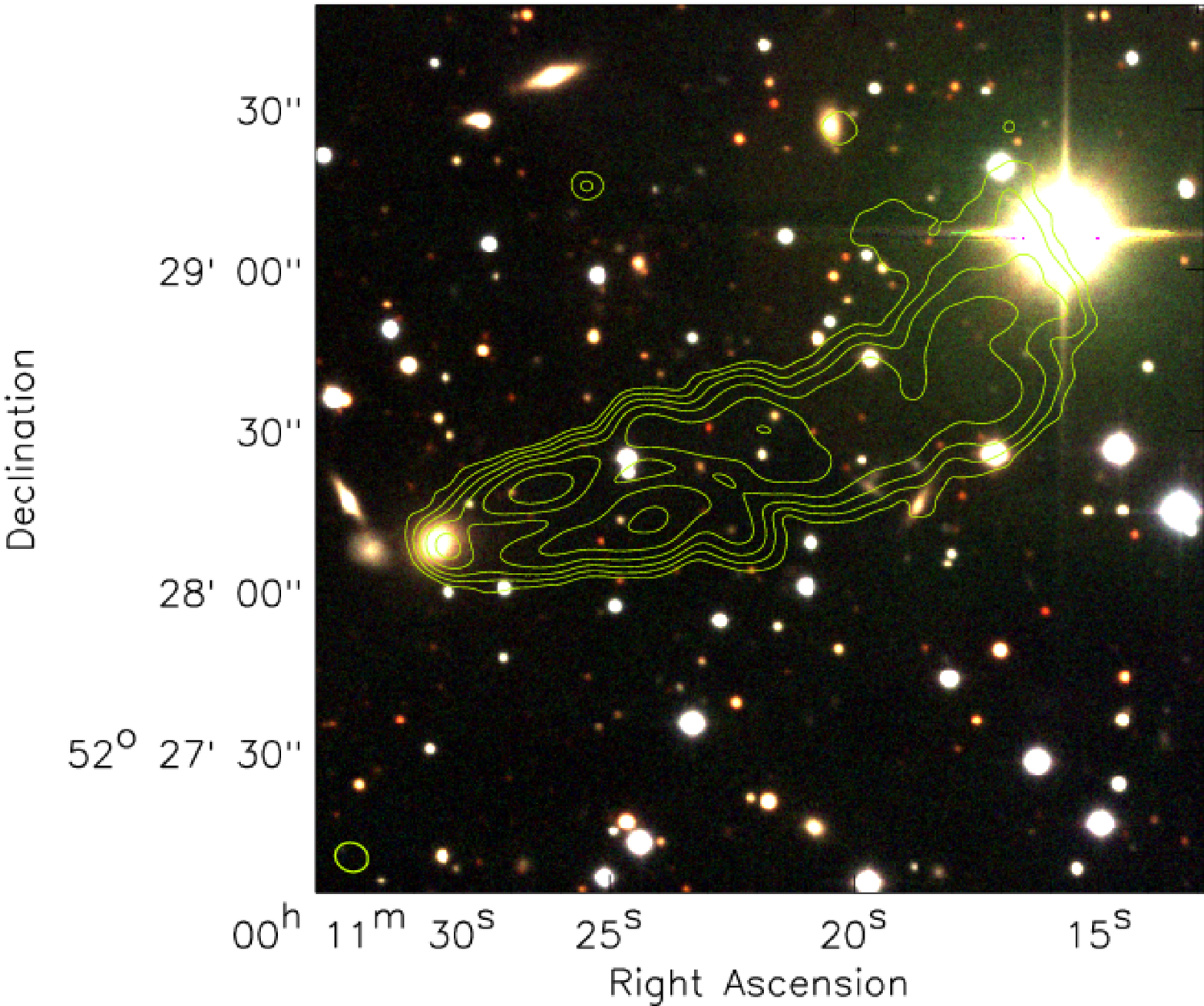}
\end{center}
\caption{Source C. GMRT 610~MHz contour levels are displayed as in Fig.~\ref{fig:SA}.}
\label{fig:SC}
\end{figure}

\begin{figure}
\begin{center}
\includegraphics[angle =90, trim =0cm 0cm 0cm 0cm,width=0.5\textwidth]{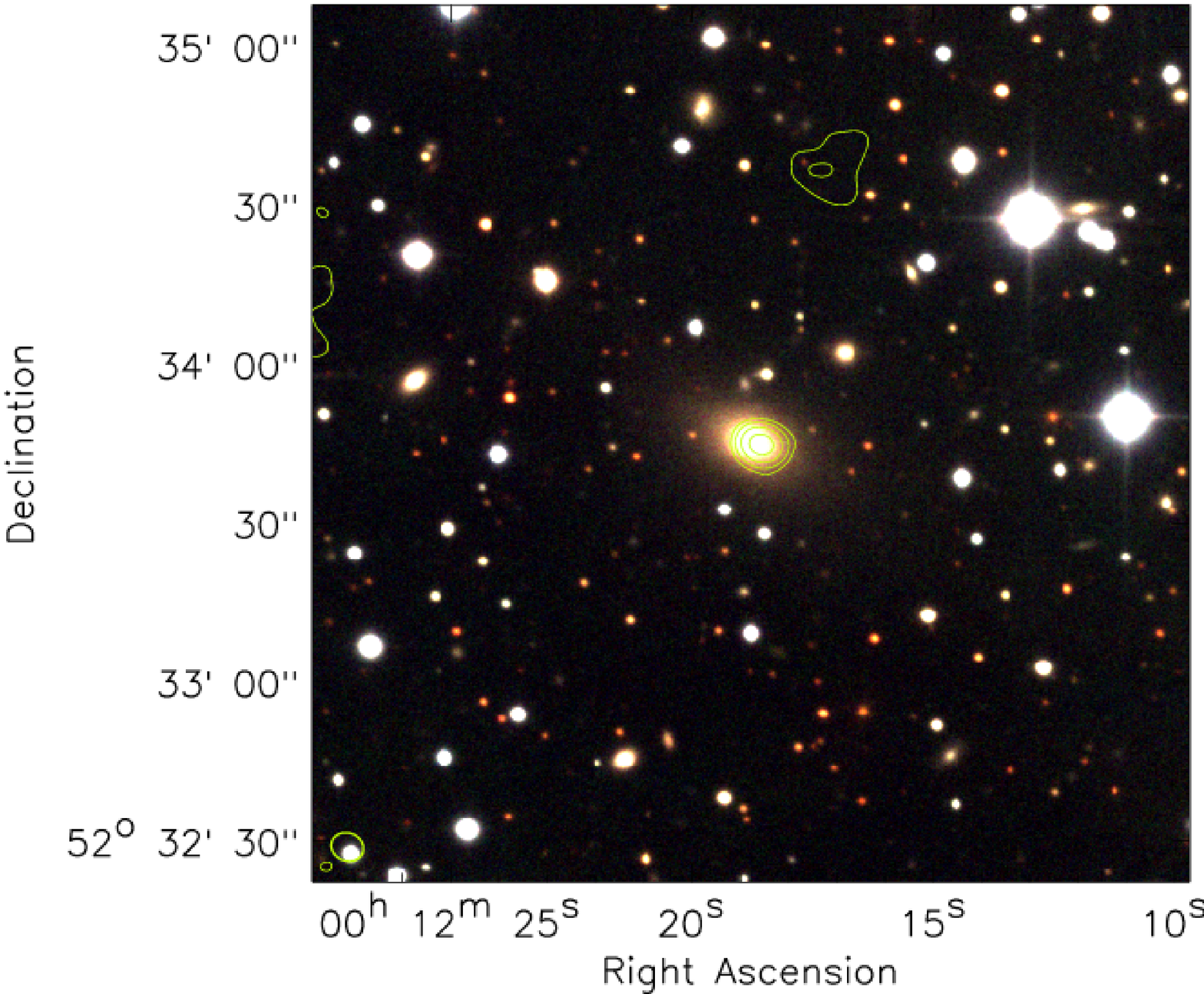}
\end{center}
\caption{Source D. GMRT 610~MHz contour levels are displayed as in Fig.~\ref{fig:SA}.}
\label{fig:SD}
\end{figure}

\begin{figure*}
\begin{center}
\includegraphics[angle =90, trim =0cm 0cm 0cm 0cm,width=0.95\textwidth]{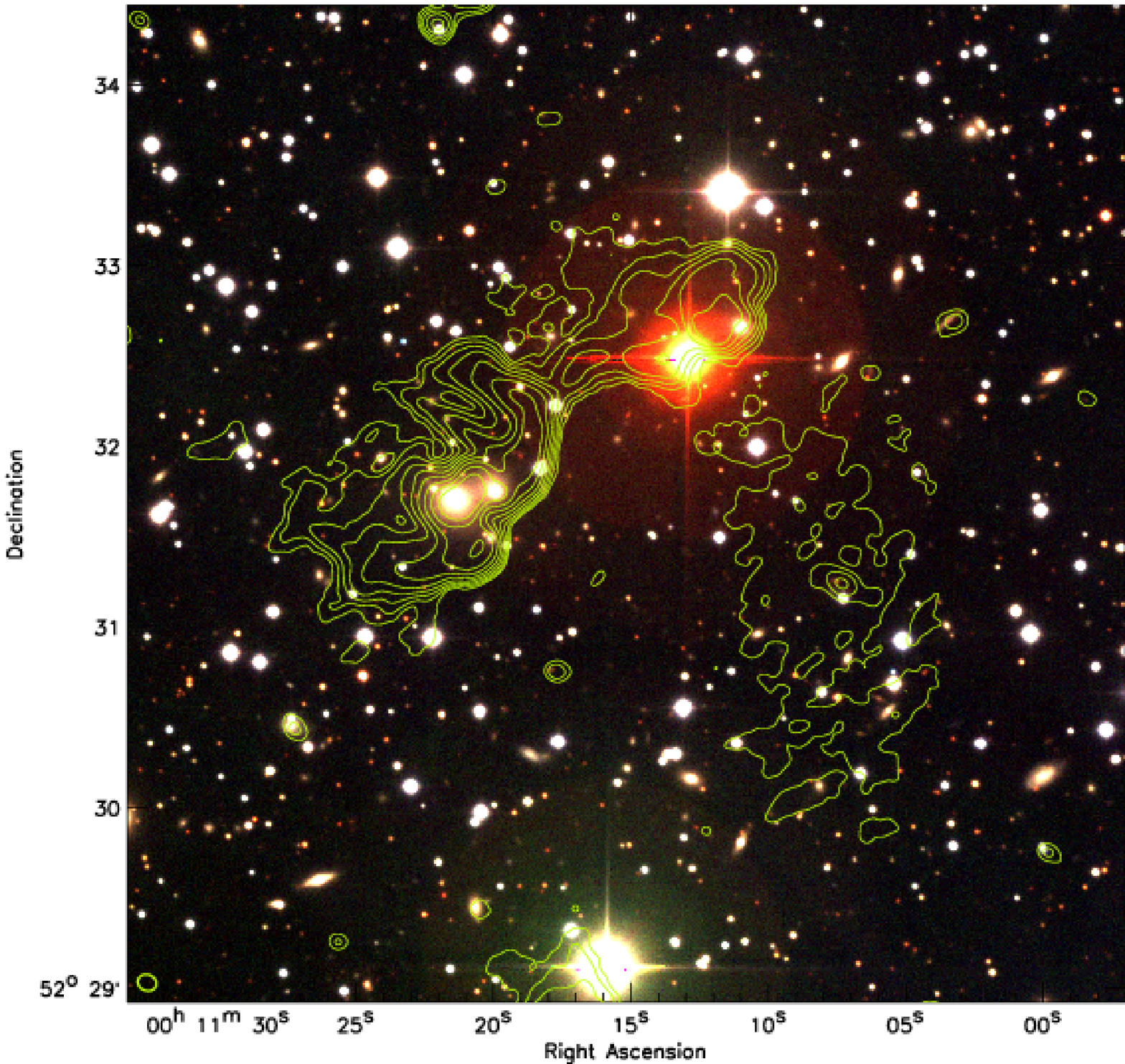}
\end{center}
\caption{Source E \& F. GMRT 610~MHz contour levels are displayed as in Fig.~\ref{fig:SA}.}
\label{fig:SEF}
\end{figure*}

\section{Discussion}
\label{sec:discussion}

\subsection{Origin of the double radio relic}

Hydrodynamical simulations of cluster mergers show that 
the process takes of the order of $10^{9}$~yr 
\citep[e.g.,][]{2010arXiv1003.5658V, 2010arXiv1001.1170P,2009MNRAS.393.1073B,2008ApJ...689.1063S, 2008MNRAS.391.1511H, 2008MNRAS.385.1211P, 2001ApJ...561..621R}. 
During a cluster merger, ``internal'' shock waves are 
generated. Typically, these have lower Mach numbers 
($\mathcal{M} \lesssim 5$) than ``external'' shock waves 
which are generated by the infall of unprocessed gas from 
the surrounding IGM and having $\mathcal{M} > 10$. The Mach 
numbers of internal shocks are low because the sound speed 
in the gas of the main (bigger) cluster and the velocity 
of the in-falling subcluster both reflect the same 
gravitational potential of the main cluster. Merger 
events which generate shock waves with $\mathcal{M} \gtrsim 3$ 
are rare, and these are mainly formed in major 
merger events, with the mass ratio of the two merging clusters 
approaching unity.  In the case of a binary cluster,  
merger two shocks are produced along the merger axis. 
As the shock waves propagate outward into a lower 
density environment their Mach number increases. The shock 
structure may get broken when it interacts with the 
galaxy filaments connected to the cluster. This may 
explain the presence of the ``notch-like'' feature 
observed in the north of relic RE \citep{2010arXiv1001.1170P}. 
These notch-like structures are also observed for 
the double relics in Abell~3376~\citep{2006Sci...314..791B} 
and Abell~3667~\citep{1997MNRAS.290..577R}.

At the location of the shock front particles are 
proposed \citep{1998A&A...332..395E} to be (re)accelerated 
by the DSA mechanism. In this case, the injection 
spectral index of the radio emission is related to the 
Mach number of the shock. Behind the shock front particles 
cool through IC and synchrotron losses. The spectral 
index should thus steepen inwards to the cluster center. 
The overall integrated spectral is still a power-law though, 
but steeper by about 0.5 unit 
\citep[e.g.,][]{1987PhR...154....1B, 1999ApJ...520..529S, 2002ASSL..272....1S} 
than the injection spectral index. Relics which such 
power-law spectra have been found for example in Abell~521 and 
the Coma cluster \citep{2008A&A...486..347G, 1991A&A...252..528G}.

We have fitted a power-law radio spectrum for the integrated 
fluxes of the radio relic reported in Table~\ref{tab:relicflux}. 
The fitted radio spectra for the two relics are displayed in 
Fig.~\ref{fig:relicflux}. RE is well fitted by a single 
power-law spectrum with a spectral index of $-1.59 \pm 0.06$.  
For the western relic we find a spectral 
index of $-1.49 \pm 0.12$. We note that the 241~MHz flux 
measurement and the corresponding error are somewhat 
difficult to estimate as the noise in the image increases 
sharply towards bright radio source E  because of residual 
calibration errors, see also Fig.~\ref{fig:gmrt241}.

As reported in Sect.~\ref{sec:spixmap}, the spectral 
indices at the front of the relics are about $-1.2 \pm 0.2$ 
and $-1.0 \pm 0.15$ for RE and RW, respectively. The errors 
give the variation in spectral index across the outer 
edges of the relic. The integrated spectral indices 
are consistent with this values (if we assume that 
DSA takes place), being steeper by about 0.5 units. 
We take the values at the front of the relic as the 
injection spectral indices. This then gives Mach numbers 
of $2.2^{+0.2}_{-0.1}$ and $2.4^{+0.4}_{-0.2}$ for 
relics RE and RW, respectively \citep{2009A&A...506.1083V}. The uncertainties in 
the Mach number are based on the variation in spectral 
index at the front of the relics of about $-0.15$ units. 
These Mach number are in agreement with those found 
in other merging clusters, typically being between 
1.5 and 3 
\citep{2002ApJ...567L..27M, 2005ApJ...627..733M, 2006ESASP.604..723M, 2010arXiv1004.1559R, 2010arXiv1004.2331F}.

Compression of fossil radio plasma  
\citep{2001A&A...366...26E} does not seem 
to be a likely scenario to explain 
the relics in ZwCl~0008.8+5215, because regions with a spectral 
index of $-0.9$ are seen in front of the relic, not 
very steep for a radio phoenix. The integrated radio 
spectra do also not reveal significant spectral 
curvature \citep[see for example][]{2001AJ....122.1172S}. 
Furthermore, relic RE has a size of 1.4~Mpc and 
the time to compress such a large radio ghost 
would have removed most of the energetic particles 
responsible for the radio emission \citep{2006AJ....131.2900C}. 
Relic RW is located not far from the complex sources E and F. If RW is directly associated with E and F, one would not expect the spectral index to steepen across RW in the direction of E and F. Also, the overall polarization fraction of RW is considerably higher than that of sources E and F.  
Therefore, our preferred scenario is that of relics tracing 
shock fronts where particles are accelerated or re-accelerated by the DSA mechanism.

The large difference of a factor of five in linear 
extent for relic RE and RW, is unlike that of 
previously known double relics. This large ratio in 
linear size could be explained by a relatively large 
mass ratio between the merging clusters, as the size of the shock waves formed 
during a binary cluster merger event roughly scale with the radii/masses of the merging components. 
The galaxy iso-density contours in Fig.~\ref{fig:xray} show two subclusters, with the eastern subcluster being slightly larger. Another possibility is that the shock front on the 
west side of the cluster is partly broken up due to 
the presence of substructures or galaxy filaments. 
In case of re-acceleration, a supply of fossil electrons is needed. Therefore, a third possibility is that 
if these electrons have a limited spatial distribution, the shock front might be illuminated only locally. 

The overall configuration of the relics 
is largely symmetric around the east-west merger axis, 
indicating that the impact parameter for the merger 
is close to zero \citep{2001ApJ...561..621R}.

\subsection{Radio luminosity profile for the eastern relic}

\begin{figure}
\begin{center}
\includegraphics[angle =90, trim =0cm 0cm 0cm 0cm,width=0.5\textwidth]{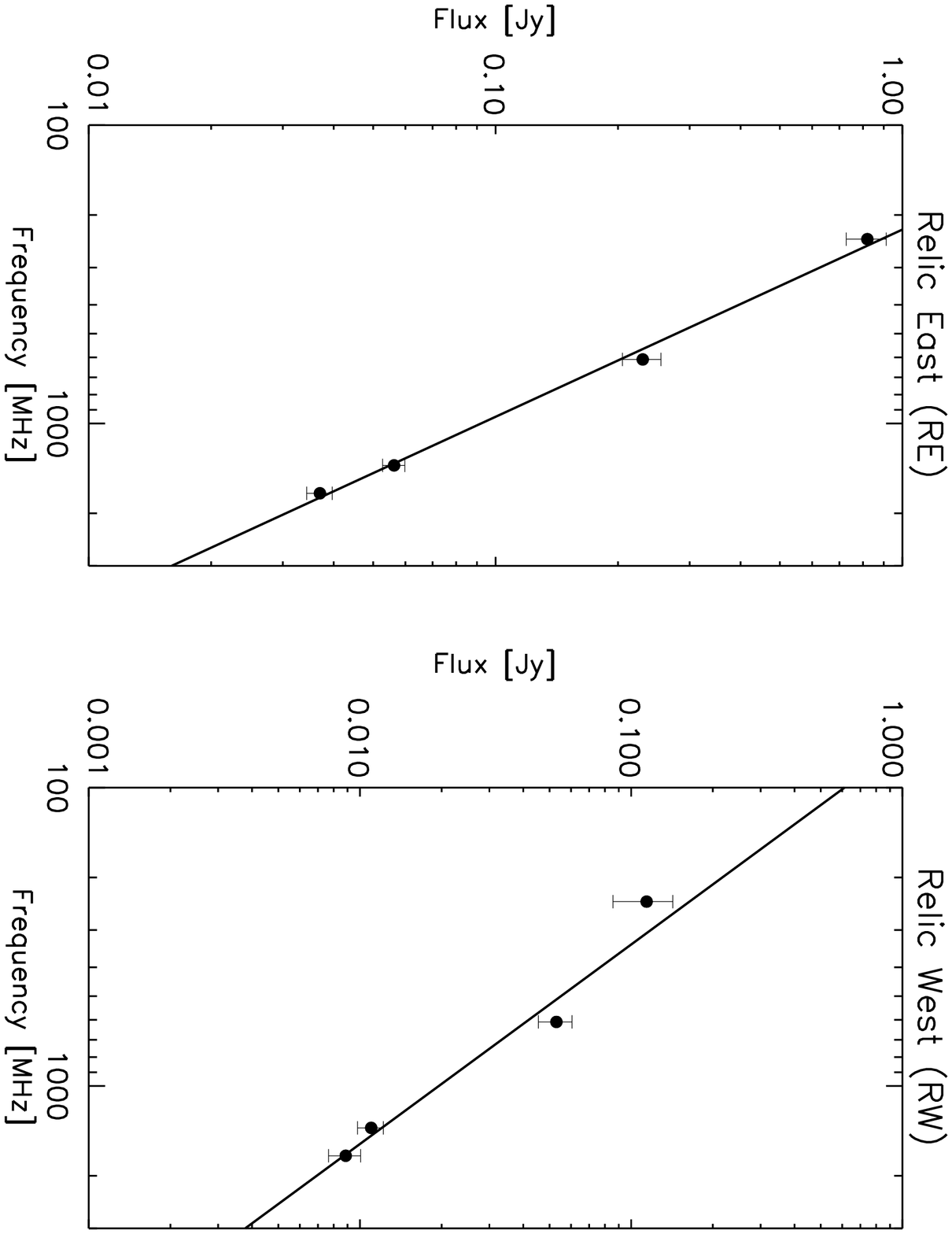}
\end{center}
\caption{Integrated fluxes for the relics. We find a 
spectral index of $-1.59$ for relic RE and $-1.49$ 
for relic RW by fitting a single power-law spectrum 
through the flux measurements at 241, 610, 1382, and 1714~MHz.}
\label{fig:relicflux}
\end{figure}

\begin{figure}
\begin{center}
\includegraphics[angle =90, trim =0cm 0cm 0cm 0cm,width=0.5\textwidth]{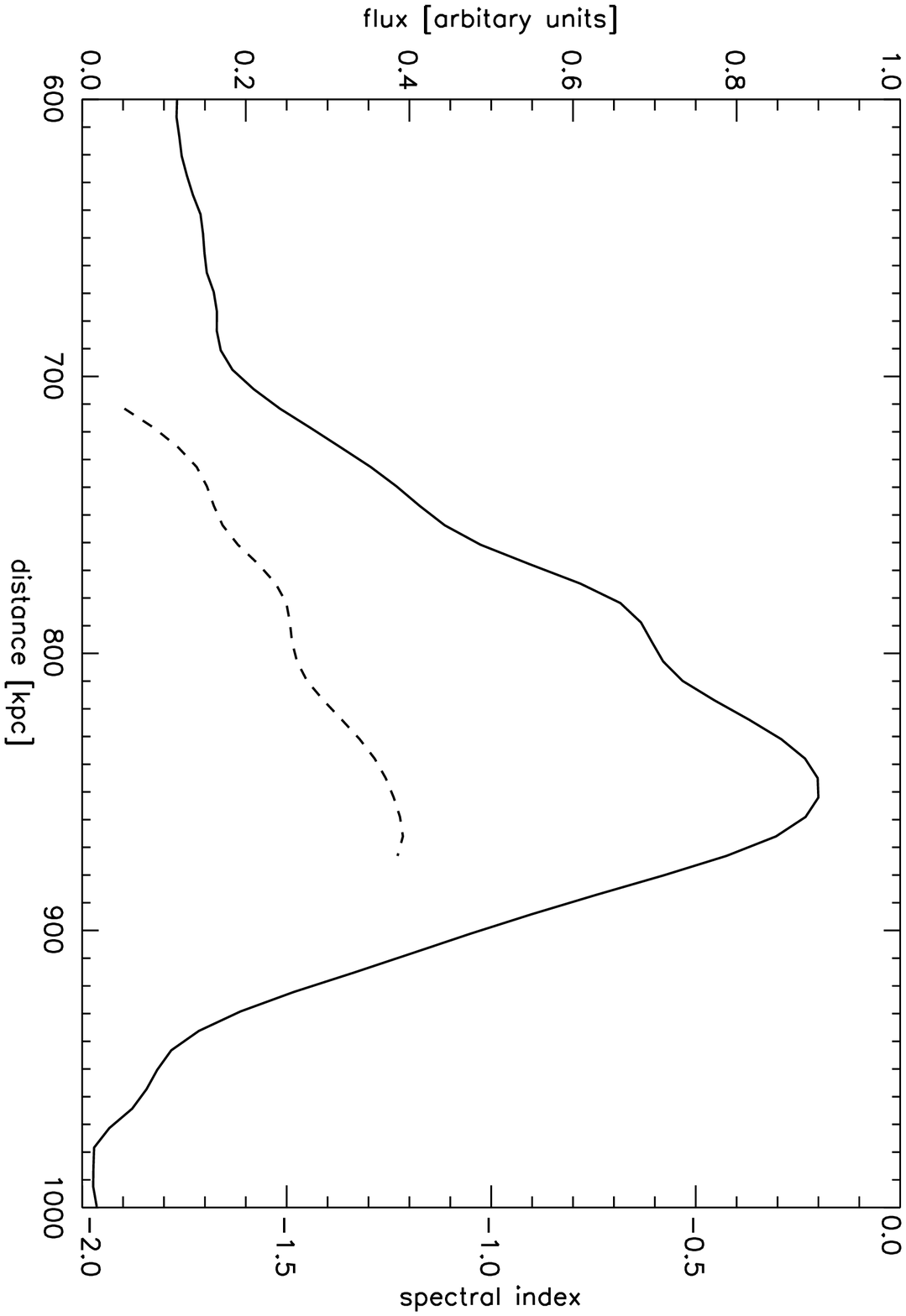}
\end{center}
\caption{Radio luminosity profile across the width 
of relic RE. Solid line displays the 1382~MHz luminosity 
profile, averaged over the full 1.4~Mpc extent, of 
the relic RE. Dashed line displays the resulting spectral 
index profile (labeling on the right axis), also 
averaged over the full 1.4~Mpc extent, between 1382 
and 241~MHz. The averaging was done by adding up 
the total flux at each frequency in a spherical 
shell and then calculating the spectral index.}
\label{fig:relicprofile}
\end{figure}

 The width of a relic ($l_{\mathrm{relic}}$), tracing a 
plane shock wave, to first order reflects the downstream 
velocity of the shock $v_2$, a characteristic timescale 
due to spectral ageing $t_{\mathrm{ageing}}$, and the 
angle $\phi$ between the shock front normal and the plane of the sky
\begin{equation}
 l_{\mathrm{relic}} \approx \frac{v_2 \times t_{\mathrm{ageing}}} { \cos{\phi}} \mbox{ .} 
\label{eq:lrelic1}
\end{equation}
The characteristic timescale due to spectral 
ageing is given by
\begin{equation}
t_{\mathrm{ageing}} \mbox{ [yr]} = 3.2 \times 10^{10} \frac{B^{1/2}}{B^2 + B^{2}_{\mathrm{CMB}}} \left[(1+z)\nu\right]^{-1/2} \mbox{ ,} 
\label{eq:tage}
\end{equation} 
with $B$ the magnetic field at the shock front in $\mu$Gauss, 
 $B_{\mathrm{CMB}}$ the equivalent magnetic 
field strength of the CMB in $\mu$Gauss, and $\nu$ the observed frequency in MHz. At $z=0.103$, 
$B_{\mathrm{CMB}}$ is $4.0~\mu$Gauss. If $v_2$ and $\phi$ 
are known this gives a method for determining the magnetic 
field strength. Even if $\phi$ is not known, limits on 
the magnetic field can be obtained if the observed 
width ($l_{\mathrm{relic}}$) is smaller than the 
maximum width allowed from Eq.~\ref{eq:lrelic1}.

To get an estimate of $v_2$, we use a temperature 
in the post-shock region of 6~keV, i.e, 
about twice the average cluster temperature. 
This factor of two increase is roughly what has 
been observed in other clusters with shocks 
\citep[e.g.,][]{2010arXiv1004.1559R, 2009ApJ...693L..56M}. 
We use Rankine-Hugoniot jump conditions 
\citep{1959flme.book.....L}, with an adiabatic 
index $\gamma =5/3$, and take the Mach number from 
the injection spectral index. This gives
 \begin{equation}
 \frac{T_2}{T_1} = \frac{5\mathcal{M}^4 + 14 \mathcal{M}^2  - 3}{16\mathcal{M}^2 } \mbox{ ,}
 \end{equation}
with indices 1 and 2 referring to the 
pre-shock and post-shock regions. 
The downstream speed is given by $ v_2= \mathcal{M} c_{s,1} / C$, with $c_{s,1}$ 
the pre-shock sound speed, 
$\left( \gamma k_{\rm{B}} T_1/ m_{\rm{H}} \mu   \right)^{1/2}$, with $\mu=0.6$ 
the mean molecular weight. The compression ratio $C$ is given by

\begin{equation}
\frac{1}{C} =   \frac{3}{4\mathcal{M}^2} + \frac{1}{4}           \mbox{ .}  
\end{equation}
 Filling in the numbers gives $C=2.4$ and $c_{s,1}=1100$~km~s$^{-1}$. 
We then obtain $v_2 = 750$~km~s$^{-1}$. The  downstream velocity 
depends only weakly on the adopted downstream temperature. For 
example, using a downstream temperature of 10~keV  
increases of $v_2$ to about $950$~km~s$^{-1}$. For the remainder 
we will adopt  $v_2 = 750$~km~s$^{-1}$. This then gives for the 
width of the relic (FWHM) observed at 1382~MHz
\begin{equation}
l_{\mathrm{relic}} \mbox{ [kpc]}   \approx  {628 \times \frac{B^{1/2}}{B^2 + B^{2}_{\mathrm{CMB}}}}{\cos^{-1}{\phi}} \mbox{ ,}
\label{eq:lrelic}
\end{equation}
 with the magnetic field strengths in units of $\mu$Gauss.
 
For $\phi=0\degr$, the maximum width is 46~kpc, 
which corresponds to $B\approx2~\mu$Gauss. This is 
 smaller than the observed width of about 150~kpc 
(see Fig.~\ref{fig:relicprofile}) and hence no 
constraints on the magnetic field can be put since the  
angle $\phi$ is not known. It is possible to set 
limits on $\phi$ using the observed polarization 
fraction \citep{1998A&A...332..395E}. A 20\% polarization 
fraction implies $\phi <50\degr$. This limit on $\phi$ 
is not consistent with the observed width which would 
require $\phi > 72\degr$. Although, for large sections 
of the relic the polarization fraction is 
unknown and could be smaller than 20\%. 

\subsection{Simulated radio luminosity and spectral index profiles }

In the above analysis we assumed that a relic traces a 
planar shock wave. In a more realistic model of a 
relic would trace a shock wave that forms a part of 
a sphere. This is   
illustrated by the curved shape of relic RE. The 
observed width is about a factor of three larger than 
the maximum intrinsic allowed width, $\max({l_{\mathrm{relic} }(\phi=0, B)})$. This implies that 
projection effects probably play an important role. 
The questions is then why do we still see a clear 
spectral index gradient (Fig.~\ref{fig:relicprofile}) 
across the relic?

To answer this questions we use a more realistic 
model of a shock front. The spherical shock subtends 
an angle $\Psi$ into the plane of the sky and has 
a radius of curvature $R_{\mathrm{projected}}$. The 
total angle subtended by the relic is $2\Psi$. We 
compute the radio luminosity profiles at the observed 
frequencies of 241 and 1382~MHz. The injection 
spectral index is taken to be $-1.0$. Synchrotron 
cooling processes, based on the distance of the 
emitting radio plasma from the front of the shock, 
which in turn depend on the downstream 
velocity $v_2 = 750$~km~s$^{-1}$, are taken into account. 
For the magnetic field we assume $B=2~\mu$Gauss, 
which maximizes the intrinsic width of the 
relic to 46~kpc. A spectral index profile 
is computed using the profiles at the two 
different frequencies. The resulting intrinsic 
luminosity profiles (with intrinsic referring 
to a planar shock wave without any projection 
effects) and profiles for $R_{\mathrm{projected}}=0.75$ and 1.0~Mpc, 
with opening angles $\Psi=22,30,40\degr$  
are shown in Figs.~\ref{fig:profileMH750} and 
\ref{fig:profileMH1000}. For $R_{\mathrm{projected}}=0.75$~Mpc, we 
find that the profile with an opening angle between 30 and 
22\degr ($\sim 26\degr$) provides the best match to the observed 
profile. For $R_{\mathrm{projected}}=1.0$~Mpc, we find the best
match for $\Psi=22\degr$.

Our computed luminosity profiles do no 
provide a very good match to the observed 
profile at distances of more than 0.85~Mpc 
from the cluster center. The observed profile 
is more symmetric, while the computed profiles 
are rather asymmetric with a strong luminosity 
decrease at large radii. This may be caused by 
the fact that the actual 3D shape of the shock 
front differs somewhat from a sphere. Also, we 
assumed a uniform surface brightness over the 
front of the shock surface (which forms part of 
a segment of a sphere). At the edges of this 
surface the radio emission drops to zero abruptly. 
This causes the discontinuities in the modeled 
profiles inwards of the peak luminosity towards 
the cluster center. 
In the GMRT and WSRT images,  
the relic's surface brightness fades  
towards the northern and southern ends. This could (partly) be 
explained by the spherical shell model we use for the relic, as the relic's extent into the plane of 
the sky decreases at the northern and southern ends. 
It is also possible that the surface brightness across 
the shell decreases towards the edges. This effect is not 
included in our model. However, our goal was not 
to reproduce the exact profile of the relic, 
but rather to show that although projection effects 
can be significant, a clear spectral index gradient 
can remain. Even for an opening angle of 40\degr  
(a total of 80\degr~into and out of the plane 
of the sky), a steepening of more than 0.5 units 
in the spectral index is predicted towards the cluster center.

Based on this we argue that although relic RE is 
widened significantly by projection effects, the fact 
the we see a clear spectral index gradient is not 
surprising. This could also explain the spectral index 
gradients visible for the relics observed by 
\cite{1997MNRAS.290..577R, 2006AJ....131.2900C, 2007A&A...467..943O, 2008A&A...486..347G, 2009A&A...494..429B}, 
even though the observed widths 
are significantly larger than the maximum intrinsic widths.

\begin{figure}
\begin{center}
\includegraphics[angle =90, trim =0cm 0cm 0cm 0cm,width=0.5\textwidth]{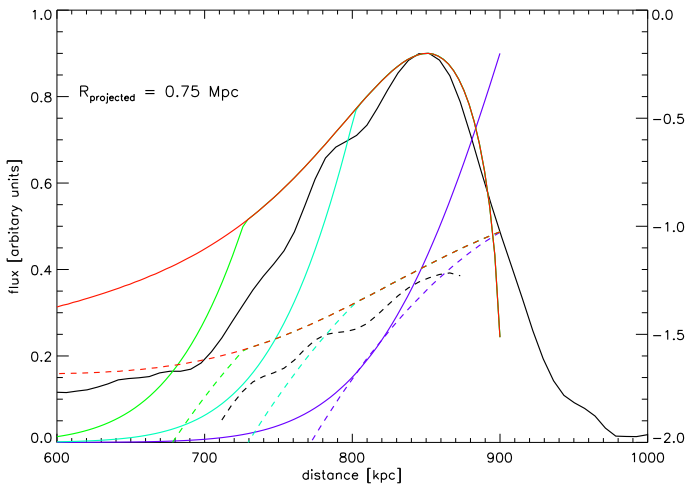}
\end{center}
\caption{Radio luminosity profile 
for $R_{\mathrm{projected}}=0.75$~Mpc and $B=2$~$\mu$Gauss. 
Solid black line shows the observed luminosity 
profile for relic RE at 1382~MHz, while the 
dashed black line represents the spectral index profile 
between 241 and 1382~MHz. The simulated intrinsic   
luminosity and spectral index profiles  (i.e., without any projection effects), are shown in 
purple by solid and dashed lines, respectively. 
The simulated profiles, including the projection effects, are 
shown by solid lines for opening angles of 22\degr (cyan), 30\degr  
(green), and 40\degr (red). The corresponding spectral index 
profiles are shown by the dashed lines.}
\label{fig:profileMH750}
\end{figure}

\begin{figure}
\begin{center}
\includegraphics[angle =90, trim =0cm 0cm 0cm 0cm,width=0.5\textwidth]{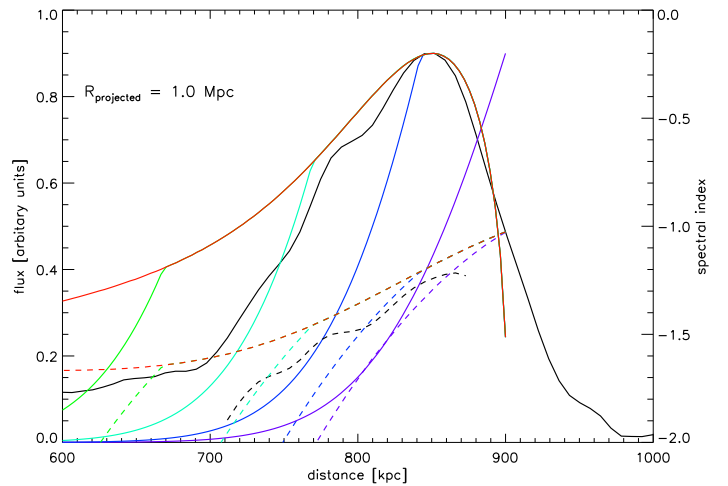}
\end{center}
\caption{Radio luminosity profile for $R_{\mathrm{projected}}=1.0$~Mpc. 
See the caption  of Fig.~\ref{fig:profileMH750} for an explanation 
of the various profiles. In addition, luminosity and spectral index 
profiles for an opening angle of 15\degr~are shown in blue.}
\label{fig:profileMH1000}
\end{figure}

\subsection{Equipartition magnetic field strength}
Since the width of the relic is larger than the maximum intrinsic width, we 
estimate the magnetic field at the location of the relics by assuming 
minimum energy densities in the relics. We use the same procedure 
as described in \cite{2009A&A...506.1083V} and take $k=100$, i.e, the 
ratio of energy in relativistic protons to that in electrons. For relic RW,  
we have a spectral index of -$1.49$, and a surface brightness of $1.2~\mu$Jy~arcsec$^{-2}$. 
We take 290~kpc for the depth ($d$) along the line of sight. This gives $B_{\mathrm{eq}} = 3.4~\mu$Gauss. For RE we have a spectral 
index of -$1.59$, a surface brightness of $0.96~\mu$Jy~arcsec$^{-2}$, and we assume 
$d=1$~Mpc. This gives $B_{\mathrm{eq}} = 2.5~\mu$Gauss. 
The equipartition magnetic field strength scales with $(1+k)^{2/7}$. In the 
above calculation, we used fixed frequency cutoffs ($\nu_{\mathrm{min}} = 10$~MHz 
and $\nu_{\mathrm{max}} = 100$~GHz), which is not entirely correct 
\citep{2005AN....326..414B, 1997A&A...325..898B}.  With low and high 
energy cutoffs ($\gamma_{\mathrm{min}}$, $\gamma_{\mathrm{max}}$), $\gamma_{\mathrm{min}} \ll  \gamma_{\mathrm{max}}$, and fixing 
$\gamma_{\mathrm{min}}$ to 100, we find a revised magnetic field strength 
($B^{\prime}_{\mathrm{eq}}$) of 7.9 and 6.6~$\mu$Gauss for RW and RE, 
respectively. For a lower cutoff of $\gamma_{\mathrm{min}}=5000$, we get 1.4  
and 1.0~$\mu$Gauss for RW and RE, respectively. The revised equipartition 
magnetic field strength ($B^{\prime}_{\mathrm{eq}}$) scales with 
$(1+k)^{1/(3-\alpha)}$, for different values of $k$.

\section{ Conclusions}
\label{sec:conclusion}
We discovered a double radio relic in the galaxy cluster ZwCl~0008.8+5215, located at $z=0.103$ (based on a single spectroscopic redshift).  
A ROSAT X-ray image and galaxy iso-density map show that the cluster is 
undergoing a binary merger event, with the merger axis oriented 
roughly east-west. The two radio relics are located along 
this merger axis, while their orientation is perpendicular to this axis. 
The relics probably trace shocks waves in the ICM, created by the 
merger event, in which particles are (re)accelerated by the DSA mechanism. 
Integrated radio spectra are consistent with particle 
acceleration in the shock by DSA and indicate Mach numbers 
of $\sim 2$ for the shocks. The spectral index for both relics
shows a steepening towards the cluster center.  
Parts of the relics have a polarization fraction in the range of $5-25\%$,   
but further observations are needed to better map the polarization properties.   
The relics have an extent of 1.4~Mpc and 290~kpc. 
This factor of five difference in their linear extent 
is unlike that of other known double relic systems. 
The size difference could be related to a relatively large mass 
ratio between the two merging clusters, although galaxy iso-density 
contours do not indicate a large difference in masses between the two 
subclusters. Alternatively, the second shock front is partly broken up due 
to interaction with substructures, or the small size of the western relic 
reflects the limited spatial distribution of fossil electrons.

We modeled the radio luminosity and spectral index 
profiles of the eastern relic, assuming that the relic 
traces a curved shock front. We conclude that projection 
effects play an important role in increasing the observed s
width of the relic. However, we find that a clear spectral 
index gradient remains visible for large opening angles. 

Future X-ray observations will be needed to 
investigate the dynamical state of the cluster, 
determine the mass ratio of the merging systems, 
and search for shock waves associated with the relics.

\begin{acknowledgements}
We would like to thank the anonymous referee for useful comments. 
We thank the staff of the GMRT who have made these 
observations possible. The GMRT is run by the National 
Centre for Radio Astrophysics of the Tata Institute of 
Fundamental Research. The Westerbork Synthesis Radio 
Telescope is operated by ASTRON (Netherlands 
Institute for Radio Astronomy) with support from the 
Netherlands Foundation for Scientific Research (NWO). 
The National Radio Astronomy Observatory is a facility 
of the National Science Foundation operated under 
cooperative agreement by Associated Universities, Inc.  
The Isaac Newton Telescope and   William Herschel Telescope are operated on the island of 
La Palma by the Isaac Newton Group in the Spanish 
Observatorio del Roque de los Muchachos of the 
Instituto de Astrof\'{\i}sica de Canarias. 
This publication makes use of data products 
from the Two Micron All Sky Survey, which is a 
joint project of the University of Massachusetts 
and the Infrared Processing and Analysis Center/California 
Institute of Technology, funded by the National Aeronautics 
and Space Administration and the National Science Foundation. 
This research has made use of the VizieR catalogue 
access tool, CDS, Strasbourg, France. 

RJvW acknowledges funding from the Royal 
Netherlands Academy of Arts and Sciences. MB acknowledges support by the research group FOR 1254 funded by the Deutsche Forschungsgemeinschaft

\end{acknowledgements}

\bibliographystyle{aa}
\bibliography{16185.bib}

\end{document}